# THE GENERALIZED LIQUID DROP MODEL ALPHA-DECAY FORMULA: PREDICTABILITY ANALYSIS AND SUPER-HEAVY ELEMENT ALPHA HALF-LIVES

N. DASGUPTA-SCHUBERT*

Instituto de Investigaciones Químico-Biológicas,
Universidad Michoacana de San Nicolás de Hidalgo, Cd. Universitaria, Morelia,
Michoacan C.P. 58060, Mexico.

and M.A. REYES

Department of Electrical Engineering, University of Texas Pan American,
1210 W. University Drive, Edinburg, TX, 78539, USA.

*Author for correspondence. Email: nita@ifm.umich.mx

The predictive accuracy of the generalized liquid drop model (GLDM) formula for alpha decay half-lives has been investigated in a detailed manner and a variant of the formula with improved coefficients is proposed. The method employs the experimental alpha half-lives of the well-known alpha standards (REFERENCE) to obtain the coefficients of the analytical formula using the experimental $Q_\alpha$ values (the DSR-E formula), as well as the finite range droplet model (FRDM) derived $Q_\alpha$ values (the FRDM-FRDM formula). The predictive accuracy of these formulae were checked against the experimental alpha half-lives of an independent set of nuclei (TEST) that span approximately the same Z,A region as the standards and possess reliable alpha spectroscopic data, and were found to yield good results for the DSR-E formula but not for the FRDM-FRDM formula. The two formulae were used to obtain the alpha half-lives of super-heavy (SHE) and heavy nuclides where the relative accuracy was found to markedly improve for the FRDM-FRDM, which corroborates the appropriateness of the FRDM masses and the GLDM prescription for high Z,A nuclides. Further improvement resulted, especially for the FRDM-FRDM formula, after a simple linear optimization over the calculated and experimental half-lives of TEST was used to re-calculate the half-lives of the SHE and heavy nuclides. The advantage of this optimization was that it required no re-calculation of the coefficients of the basic DSR-E or FRDM-FRDM formulae. The half-lives for 324 medium-mass to super-heavy alpha decaying nuclides, calculated using these formulae and the comparison with experimental half-lives, are presented.

# CONTENTS

1. **Introduction**

A number of recent advances in the synthesis of very heavy and superheavy elements (SHE) [1,2] as well as highly neutron deficient medium mass nuclei [3], have spurred theoretical investigations into the decay modes and masses of these nuclei. These investigations aim to test and advance ideas on the extant nuclear structure for heavy and high Coulomb repulsive force containing nuclei. From the stand-point of heavy element chemistry, the isolation and the study of the properties of exotic heavy elements and their compounds serve to provide unique insights into the structure, bonding and reaction energies of massive multiply charged atomic and molecular species [4]

Often the main and sometimes the only decay mode of these nuclei is alpha decay. For the SHE, the product nuclides of hot-fusion heavy ion reactions closest to the shell model predicted Island of Stability, decay by the emission of alpha particles. Chains of successive alpha decays are terminated by spontaneous fission as the shell-stabilized region is left behind [5, 2].Thus identifying and characterizing the alpha decay sequences form a crucial part of the identification of SHE. Theoretically the mechanism is described by quantum mechanical tunneling through the potential energy barrier leading from the mother nucleus to the daughter nucleus and alpha particle. Consequently the predicted half-lives remain very sensitive to the shape and energetics of the barrier which as such, serve to test the particular theoretical model of the potential energy surfaces of these exotic nuclei. Various theoretical prescriptions since 1930 [6] have been proposed. Some of these as well as some empirical observations have been reduced to analytical formulae [7-12] that connect $Q_\alpha$, and the Z,A of the parent nuclide and wherein coefficients are typically obtained from fits to known alpha half-lives. The underlying model's description of the potential barrier lies implicit in the formula's functional form and its coefficients and this in part determines the merit of the formula.

For the experimentalist, the analytical formula is especially useful because it permits easy incorporation into experimental data processing/analysis systems as well as the possibility of the upgrade of the coefficients of the formula as the experimental data base expands and/or gets refined. Ken Moody representing the Dubna-Livermore collaboration on SHE synthesis [2], mentions their use of the Geiger-Nutall relationship [7] to check their assignment of the Z of the composite system to the parent nuclides of each alpha decay chain. The experimental investigation of these exotic alpha emitters is constrained by the smallness of the number of atoms formed, the shortness of half-lives as well as the varying degrees of uncertainty in the theoretical prescriptions and formulae [13].

A systematic analysis of the predictive accuracy of several well-known or recent analytical formulae has to date received relatively insufficient attention. Often the degree of closeness of match between the predicted half-lives and the experimental half-lives of newly discovered exotic alpha emitters is taken to be a measure of the goodness of the formula towards the prediction of unknown half-lives [14, 15]. Such an accuracy test is compromised because the experimental data of the exotics are subject to substantial uncertainties, because the coefficients of the formula have been arrived at by a fitting procedure that involves the experimental data of a large number of alpha emitters many of whose mass and spectroscopic data are not sufficiently well-known and because for many if not most of the exotics, the $Q_\alpha$ have to be taken from systematics or theoretical tabulations which have their inherent errors for the particular mass range under consideration [18]. This study is an attempt to investigate in a systematic and detailed manner the reliability of the recent alpha decay analytical formula of Royer et al [14] in terms of its predictive accuracy. Concomitantly, it devises a method for the improvement of the reliability of the formula's coefficients and explores a simple optimization procedure for the application of the resulting formula to the half-lives of heavy and SHE alpha emitters.

## 2. Method

### 2.1 The Generalized Liquid Drop Model Formula for Alpha Decay Half-Lives

Recently Royer et al [14-17] have described the potential barrier in alpha decay in terms of a quasi-molecular shape path within a Generalized Liquid Drop Model (GLDM) including the proximity effects between nucleons in a neck, and the shell effects given by the Droplet Model. The difference between the experimental $Q_\alpha$ and the GLDM calculated one was empirically corrected by adding it to the macroscopic potential energy of the parent nucleus with a linear attenuation factor that vanished at the contact point of the nascent fragments. The half-lives were deduced from the WKB barrier penetration probability as for spontaneous asymmetric fission. The theoretically calculated logarithms of half-lives for known alpha emitters wherein the experimental $Q_\alpha$ were used, agree well with the logarithms of the experimental half-lives, with a total root mean square deviation (RMSD) of 0.63, which was taken as evidence of the essential soundness of the model. However, for the subset for the heavy nuclides ($Z \geq 100$) whose

$Q_\alpha$ are estimated from systematics, the agreement is comparatively worse with an RMSD of 1.16 for the logarithmic half-lives.

Further, the model expressions involving Z, A and $Q_\alpha$ of the parent were fitted against the experimental half-lives for 373 alpha emitters to arrive at an analytical formula (eqn. 1) that for even-even (e-e), even-odd (e-o), odd-even (o-e) and odd-odd (o-o) parent nuclei, differs only in the values of the coefficients a, b and c.

$$\log_{10}[T_{1/2}(s)] = a + b.A^{1/6}.Z^{1/2} + c.Z/ Q_\alpha^{1/2} \qquad (1)$$

where $T_{1/2}(s)$ is the partial alpha half-life in seconds. The values of the coefficients a, b and c for e-e, e-o, o-e and o-o parent nuclides respectively were, (-25.31, -1.1629, 1.5864); (-26.65, -1.0859, 1.5848); (-25.68, -1.1423, 1.592); and (-29.48, -1.113, 1.6971).

In this work we are concerned only with the reduction of Royer's theoretical calculations to the analytical formula, above. While the large number of 373 nuclides is beneficial for the statistics of the fitting procedure, it is not possible to get a clear idea of the closeness of match between the formula calculated half-lives and the experimental ones because any reasonably large set of well-studied alpha emitters will have members already included in the set of 373 nuclides and will therefore not be independent. Moreover for a sizable number of these 373 emitters particularly those at the extrema of the (Z,A) scale, the alpha decay properties possess significant uncertainties and for the heaviest nuclides the $Q_\alpha$ values are those reported only from systematics. These uncertainties will affect the values of the constants evaluated for the analytical formula. Slight variations in the coefficients in eqn. 1 as also the $Q_\alpha$, will have a relatively large effect on the calculated half-lives because of the logarithmic relationship. Most of these nuclides with uncertain nuclear data in [14], have undetermined or poorly determined alpha branching fractions (e.g. $^{110,111}$Xe, $^{111}$I and $^{189}$Bi) while most of the heavy nuclides with Z > 100 have $Q_\alpha$ surmised by systematics.

## 2.2 Construction of the Method for the Predictability Test

We use the term "predictability" to connote the extent of accuracy of the GLDM formula in its prediction of the half-lives of an independent set of nuclei, i.e. nuclei whose data have not been used to arrive at the values of the coefficients a, b and c. This test of accuracy necessitates the experimental nuclear data of this independent set of nuclides, the 'TEST data' set, to be also well known. The purpose of this work is to test the predictability of the GLDM formula over a defined Z,A range using such a method with the coefficients of the formula having been arrived at using accurate and highly reliable data of a set of "REFERENCE" nuclei that cover approximately the same Z,A range.

Towards this objective we first establish a set of standard nuclei whose alpha decay properties are very well known. These, called REFERENCE nuclides, are taken to be the set of nuclides recommended as alpha energy and intensity standards for the calibration of alpha particle measurements [18a]. The experimental $Q_\alpha$ (MeV), total half-life in seconds (T) and the percentage alpha branching fraction "a", along with their respective errors were obtained from the Table of Isotopes 8$^{th}$ Edn. [18]. From these data, the alpha partial half-life in seconds, T(E), and its error obtained by propagation over the errors for T and a, were extracted. The data for the 77 REFERENCE e-e, e-o, o-e and o-o nuclides are shown in table 1. For the latter 2 sets, particularly for the o-o set, the number had to be augmented by additional nuclides because of the paucity of alpha standards in this class. These nuclides were chosen [from 18] to be those that also possess well-known spectroscopic data, the *ad hoc* indicator of which was taken to be the criterion that the nuclide possess at least 5 references on the determination of its spectroscopic data.

The coefficients a, b and c of eqn 1 were arrived at after a multi-variable regression fit [19] using the T(E), $Q_\alpha$ and (Z,A) of the REFERENCE nuclides. The magnitude of statistical precision error is expected to be higher than those calculated from a much larger data set even one with some inaccurate elements. This is an inevitable constraint as it is not possible to get an extremely large highly reliable set of data. However, this work is not concerned with the establishment of a more precise formula but rather one whose underlying data confer a high degree of confidence and which thereby can serve as "calibration formula" for the set of TEST nuclides. The coefficients a, b and c for e-e, e-o, o-e and o-o parent nuclides were obtained as

(a= -22.2671, b= -1.1908, c=1.5229); (a= -27.7933, b = -1.0031, c= 1.5813);

(a= -28.8157, b= -1.0375, c= 1.6135) and (a= -28.2982, b= -1.3555, c= 1.8265) respectively. For the purpose of distinguishability, the GLDM formula (eqn 1) using these coefficients is termed the DSR formula whereas the GLDM formula with coefficients as derived in [14] (cf section 2.1) is termed the R formula.

The TEST data set, were taken to be nuclides that covered a (Z,A) range close to the REFERENCE nuclides to reduce the possibility of the influence of Z,A dependent differences in the underlying nuclear parameters in the data analysis. These nuclides, labeled as TEST, while not alpha particle measurement standards, were nonetheless chosen to be those with well-known experimental data [18]. The criterion of selection was that the nuclide possess at least 3 references in [18] for the experimental determination of its spectroscopic data. The commonality of criteria and the general equivalence in quality between the two sets, REFERENCE and TEST, is ensured with these two indices of characterization, i.e. spectroscopic data that are well-known and a (Z,A) range that is nearly the same. The T(E) values were obtained as for the REFERENCE nuclides. The half-lives, T(DSR-E), were those obtained using the DSR formula and experimental $Q_\alpha$ values.

Moreover, for exotic nuclei, experimental $Q_\alpha$ values are generally non-existent or contain large uncertainties because of which it becomes necessary to use calculated masses appropriate to the particular Z,A region. In this work we use the Finite Range Droplet Model (FRDM) [20] for the analysis of the SHE, as it is more appropriate for heavy nuclei. The Thomas-Fermi model [21] used for the SHE in [14] and the FRDM are similar in their use the liquid drop model for the macroscopic potential energy but differ in their microscopic part. While both yield similar RMSD values over nearly equal sized Z,A ranges, the RMSD for the FRDM is only 0.448 MeV for nuclei with N≥ 65. Moreover [20] provides a comprehensive data table over the entire Z,A range.

In order to bench-mark the GLDM formula for the cases where theoretically derived $Q_\alpha$ have to be used, the coefficients a, b and c for eqn 1 were additionally derived using the FRDM $Q_\alpha$ and the T(E) values of the REFERENCE nuclides through the regression fitting process [19]. These values for e-e, e-o, o-e and o-o nuclides were respectively,

(a= -7.26534, b= -1.21302, c= 1.1441); (a= 5.59974, b= -2.04514, c=1.36734);

(a= -6.52469, b= -1.54008, c= 1.35284) and (a= -29.0028, b= -0.721707, c= 1.45336).

The half-lives for the TEST nuclides, T(FRDM-FRDM) were calculated from eqn 1 using these coefficients and the FRDM $Q_\alpha$. Additionally, the TEST nuclide half-lives were also calculated using the standard DSR and R coefficients and the FRDM $Q_\alpha$, labeled as T(DSR-FRDM) and T(R-FRDM).

As measure of accuracy we take the square of the relative error or fractional deviation, FDsq, defined as,

$$FDsq = [\{T(E) – T(Calc\text{-}x)\} /  T(E)]^2 \qquad (2)$$

Where, "Calc" stands for the mode of calculation used to arrive at the half-life: T for theory derived, DSR and R for calculations using the DSR and R analytical formulae respectively and FRDM for the GLDM formula with coefficients derived using FRDM $Q_\alpha$. The x stands for the type of $Q_\alpha$ inserted into the relevant formula or the theory to obtain the half-life: x = E stands for experimentally derived $Q_\alpha$ and x = FRDM for the FRDM calculated $Q_\alpha$. These combinations show the various calculations performed and comparisons made for the predictability analysis. For the experimental data alone, the FDsq was simply the square of the ratio of the experimental error of T(E) to the T(E) value.

### 3 Results and Discussion

#### 3.1 Analysis of Predictability of the GLDM Formula

Table 1 shows the experimental data and results of Royer's theoretical calculations [14] labeled as T(T-E), for the REFERENCE nuclides. As expected for nuclides designated as standards, the root mean fractional deviation squared, RMFDsq values for the experimental half-lives (T(E)), RMFDsq(E), are low – the least for the e-o set and relatively the most for the o-o set. The RMFDsq values for the experimental $Q_\alpha$ were negligible. In the absence of formula error, the RMFdsq(E) give the idea of the minimum uncertainty in the coefficient values that might be expected. The residual deviations of the half-lives predicted by the GLDM theory when the dominant $Q_\alpha$ component is not of concern, (since it is constructed to be equivalent to the experimental $Q_a$ values), is the root mean fractional deviation squared value for the theoretical half-lives, RMFDsq(T-E) and its ratio to the RMFDsq(E) value. On average, this latter is 8.26. This is low given the orders of magnitude spread in the half-life values. It substantiates the essential validity of Royer et al's theoretical model [14-17] and thereby the validity of the GLDM formula.

The experimental data and calculated values for the half-lives of the 225 TEST nuclides using the GLDM analytical formulae with experimental and FRDM $Q_\alpha$, i.e. T(DSR-E) and T(FRDM-FRDM), as well as the T(T-E) are shown in table 2. The R formula calculations have been omitted because most of the TEST nuclides are included in the set of 373 nuclides from which coefficients of the R formula had been derived in [14].

The root mean squared relative accuracy of the DSR formula using experimental $Q_\alpha$, RMFDsq(DSR-E), are all in the region of units with the highest value for the e-o nuclides and the least for the o-e nuclides. Under the assumption of the validity of the GLDM formula, the T(DSR-E) values are influenced by the values of the coefficients in the DSR formula and the exactness of the $Q_\alpha$ values. The experimental $Q_\alpha$ values of the TEST nuclides had negligible uncertainties. The Z,A region of the TEST nuclides is similar to the Z,A region of the REFERENCE nuclides using whose data the coefficients have been derived. Implicit in the values of the $Q_\alpha$ and T(E) of the REFERENCE are the nuclear parameters involved in the potential barrier and energetics of the mother and daughter nucleus and therefore the averaged coefficients will bear their imprints. It is expected that these averaged coefficients will apply reasonably well for the TEST nuclides of the similar Z,A range. However because of the finite differences in the nuclear parameters between the TEST and REFERENCE nuclides and because the process of fitting presents only the smoothed coefficient values, the resulting small differences produce finite deviations from the true half-life values, T(E). These differences are likely to be larger if the coefficients were obtained from a REFERENCE set with a markedly different Z,A. This method of separating the REFERENCE and TEST nuclei is thus useful in probing the general systematic variations of the relative accuracy of the GLDM formula between different Z,A regions.

The RMFDsq(T-E) values are lower because the experimental $Q_\alpha$ and individual nuclear parameters are explicitly taken into account in the calculation for each nuclide. Nonetheless, it will be observed that the RMFDsq(DSR-E) are only around a factor of 2 higher than it. This factor as well as the magnitude of the RMFDsq(DSR-E) indicate a relatively low error in the calculations in the context of the orders of magnitude variation of the half-lives of the 225 nuclides. In order to locate particular nuclide outliers, it would be interesting to observe the variation of FDSq with the Z and N numbers of the parent. These are shown in figures 1 and 2. The log(FDsq(T-E)) values span a range of mostly -3 to +1 whereas in the case of the log(FDsq(DSR-E)), the translation to the analytical formula and the least squares fitting process widens the range of the relative error to between -4 and +2. In fig.1 larger errors in the DSR-E calculations are obtained for even Z particularly for Z= 72, 74,78 and Z= 86, 92, 98 while in fig 2 the bi-modal distribution in errors appears more clearly with the larger errors being concentrated around N=87, 89, 107 and N=131, 137, 147. This accounts for the largest RMFDsq(DSR-E) values for the e-o set. However a very distinct outlier in both figures is the o-o $^{176}_{77}$Ir. If it is included in table 2, it drives up the RMFDsq(DSR-E) value for the o-o set 32 fold. The fact that such obvious outliers are restricted to only 1 over the entire set of 225 nuclides in TEST, lends support to the DSR-E formula. This nuclide has been omitted in the T(T-E) calculation of [14]. The bi-modal distribution is probably simply an artifact of the least squares fitting process and is not fundamental to the GLDM for it does not show up for the distributions of log(FDsq(T-E)). For low Z,N (medium mass alpha emitters) the T-E calculations appear to have more relative error which might indicate the greater validity of the GLDM model for heavier nuclides.

The analytical formula finds particular application in the predictions of half-lives of exotic nuclei whose $Q_\alpha$ are usually not reliably known. In the following we address the question of the quality of predictability of the GLDM analytical formula when theoretical masses have to be used. Royer [14] uses the Thomas-Fermi model derived $Q_\alpha$ values in the R formula to obtain the half-lives of SHE. For some SHE the calculated half-lives were within the range of uncertainties, albeit quite large, of the experimental data extant at that time while they were off by several orders of magnitude for the e-o SHE. Table 2 shows the FRDM $Q_\alpha$ and the half-lives of the TEST nuclides obtained using the DSR formula and FRDM $Q_\alpha$, T(DSR-FRDM), and using the FRDM-FRDM formula (coefficients of the GLDM analytical formula derived using FRDM $Q_\alpha$) and FRDM $Q_\alpha$, T(FRDM-FRDM).

The RMFDsq(DSR-FRDM) is very large for all the nucleon pairing sets (only the results for e-e are shown). The difference between the FRDM $Q_\alpha$ and experimental $Q_\alpha$ is small - for the e-e set the root mean square of the fractional deviation, RMFDsq($Q_\alpha$ FRDM) is 0.09. However, the large relative error results because of the sensitivity of the formula half-life to slight variations in the $Q_\alpha$ value and also possibly due to the lack of consistency of the type of $Q_\alpha$ used between the obtainment of the coefficients and the calculation of the half-lives. To investigate this point, the half-lives were calculated using the formula with coefficients derived using the FRDM $Q_\alpha$ for the REFERENCE nuclides. These are the T(FRDM-FRDM) values in table 2. The RMFDsq(FRDM-FRDM) was still high but got reduced by several orders of magnitude over the values for the RMFDsq(DSR-FRDM). This points to the necessity for consistency in the source of the parameter when using the analytical formula. The values of the RMFDsq(FRDM-FRDM) were again the highest for the e-o set but lowest for the o-o set.

### 3.2 The GLDM Formula and the Half-Lives of SHE

The DSR-E formula's coefficients have been derived from reliable data (the REFERENCE nuclides) and therefore are 'constant' in the sense that there will be little likelihood of the need for occasional data-base dependent upgrades. On the other hand, the R formula's coefficients have been obtained from a less reliable but wider Z,A spanning database which results in a smoothing over of the effects due to the variation of the nuclear parameters across this Z,A range. The relative predictive accuracies of the 2 prescriptions are examined via the calculated half-lives for the heavy nuclides with $Z \geq 99$ and recently discovered SHE [2]. Table 3 lists the experimental data and half-lives calculated by the DSR-E, R, FRDM-FRDM as well as the optimized DSR-E and FRDM-FRDM expressions discussed below. The data for the e-e and e-o SHE with $Z \geq 110$ were taken from [2], the heavy nuclides from [18] and Uuu($^{272}$111) from [14]. The errors on the T(E) values could not be ascertained as for many nuclides in the data base, the full spectroscopic data are not known. Also for many, the $Q_\alpha$ have been reported only from systematics. Thus the root mean fractional deviation squared (RMFDsq) values serve as approximate guides only to the relative accuracy.

Table 3 shows that the RMFDsq(DSR-E) values follow the same trend for the 4 cases of Z and N combinations as for the TEST nuclides. The absolute magnitudes (except for the e-o case) are lower than the ones for the TEST nuclides, further supporting the observation that the GLDM model may be more suitable for the high Z,A region. The RMFDsq(R-E), corresponding R-E calculations, are roughly about the same and follow the same trend. The highest value is again obtained for the e-o set, although this is about an order of magnitude lower than in the DSR-E calculation. The seemingly 'better' agreement of the R-E calculations are likely due to the fact that it is not a strictly independent calculation as most of the nuclides with $Z \leq 110$ were used in [14] to obtain the coefficients. From that stand-point, the closeness of the RMFDsq(DSR-E) values with the RMFDsq(R-E) values and their low magnitudes (except for the e-o set) lend substance to the essential underpinning of the DSR-E calculation i.e. the coefficients derived from a highly reliable but smaller data base, yield the GLDM analytical formula that has on average, good predictive accuracy.

To observe the relative accuracy of the formula for the case where no experimental data are available, we turn to the RMFDsq(FRDM-FRDM) values. The values are strikingly lower than the equivalent values for the TEST nuclides except for the o-o set for which they remain approximately the same. The FRDM model is more suitable for the heavy nuclides [22] and this together with the GLDM model's better applicability to the high Z,A region may account for the reduced values. The trend however, is nearly opposite to that obtained for the TEST nuclides, with the o-o set possessing the least relative accuracy. Reduction in magnitude notwithstanding, the values are still high in the context of good predictive accuracy of the half-lives of little known or hitherto undiscovered nuclei. We explore a possible way to optimize the FRDM-FRDM calculations while adhering to the principle of coefficient derivation from the REFERENCE data base only.

While a relatively larger statistical error is embedded in the REFERENCE's narrower data base, we attempt to 'expand' the optimization set by a linear fit of the log(T(FRDM-FRDM)) values for the TEST nuclides to their log(T(E)) values. The graphs and the linear fit equations are shown in figs 3-6. The half-lives for the heavy nuclides and SHE were then calculated using the linearly fitted relationships shown in the graphs and under Explanation of Table and Figures. The resulting half-lives, T(FRDM-FRDM-fit) are shown in table 3 along with the RMFDsq(FRDM-FRDM-fit) values. There is considerable reduction in the values of these deviations resulting in a higher degree of confidence on the accuracy of the predicted half-lives. The relative improvement is the highest for the e-e set and the least for the o-o. This improvement does not appear to be "correlated" with the magnitude of the random errors of the fits (as indicated by the values of the correlation coefficients).

A similar attempt to optimize the DSR-E calculations, was made. The results are shown in figs 3-6 and the optimized half-lives, T(DSR-E-fit) in table 3. The RMFDsq(DSR-E-fit) values while lower, are less dramatically so than the

RMFdsq(FRDM-FRDM-fit) values. There is about a factor of slightly more than 2 improvement for the e-e and e-o set whereas for the o-e and o-o set the values remain nearly the same. The degree of improvement for these calculations appears to be loosely "correlated" with the magnitude of the random error of the optimization – the linear fit with the highest correlation coefficient gives the best improvement. The value for the e-o set while lower than before, remains high. It is difficult to say why the e-o set is so distinctly different in both the DSR-E and R-E calculations. Royer [14] observed the same for the R calculations using the Thomas-Fermi model $Q_\alpha$. Nonetheless in general, the secondary optimization results in a further enhancement of the predictive accuracy of the DSR-E formula.

The improved accuracies for the half-lives of the SHE and heavy nuclides in both the FRDM-FRDM-fit and DSR-E-fit calculations are possibly also contributed by the nearness of the (Z,A) region of many nuclides in the optimization set (TEST) to the (Z,A) of the SHE and heavy nuclides so that (Z,A) dependences of the nuclear parameters have probably reduced effects. The linear fits for the DSR-E calculations in figs3-6 evince less scatter for all four sets as compared to the fits for the FRDM-FRDM calculations. The larger scatter for the latter is an indication of the errors implicit in the FRDM masses in this Z,A region although the e-o set´s largest scatter (fig 4) might also be contributed by the larger GLDM model errors for this nucleon parity set.

4. **Conclusion**

The accuracy of the generalized liquid drop model, GLDM, of Royer et al [14-17] towards the prediction of alpha half-lives, has been tested in a detailed manner. The analytical formula of [14] was modified resulting in the DSR-E formula that has the advantage of possessing coefficients derived from a reliable data base (REFERENCE). Using this formula, the predictive accuracy of the GLDM formula was checked against the experimental half-lives of a set of nuclides whose alpha decay data are well-known (the TEST) and was found to be satisfactory. The coefficients of the formula based on theoretical $Q_\alpha$, the FRDM-FRDM formula, were also obtained using the REFERENCE nuclides, where FRDM is the finite range droplet model formalism of nuclear masses [20]. Its predictive accuracy, checked in a similar manner, was found to be unsatisfactory. The DSR-E and FRDM-FRDM formulae were used to calculate the half-lives of experimentally known heavy nuclides and SHE. The DSR-E calculations produced slightly better agreement between the experimental and calculated half-lives for this set than for the TEST nuclides (except for the e-o set), probably indicating the better applicability of the GLDM prescription to the high Z,A region. The FRDM-FRDM calculations showed markedly improved calculated half-lives but still did not produce sufficiently reliable results.  A simple optimization procedure whereby the FRDM-FRDM as well as the DSR-E calculations were linearly optimized in a secondary step using the calculated and experimental data for the TEST, was explored. This modification yielded the best agreement between the calculated and experimental half-lives for the heavy nuclides and SHE, for both the FRDM-FRDM and the DSR-E although for the latter, the e-o set continued to be deviant.

The results of this work suggest that (1) the ansatz of selecting 2 data sets with well-known spectroscopic data, one for calibration and the other for test, is a valid method for the check of accuracy of a theoretical prescription for the decay half-life (2) the GLDM analytical formulae with coefficients derived from a limited but highly accurate data set, result in good agreements between model predicted and experimental alpha decay half-lives; (3) that the use of model derived $Q_\alpha$ necessitates derivation of the coefficients of the alpha decay formula using the same model for the $Q_\alpha$ for reasons of consistency; (4) that the process of simple linear optimization of the two variants of the GLDM formula using experimental or model derived $Q_\alpha$, result in the best predictive accuracies for heavy nuclides and SHE and finally (5) that the GLDM formula with FRDM masses produce better predictive accuracy for the highest Z,A alpha emitters.

Furthermore, this method of calibrating the calculation against a 'standard' data set, allows the comparison between two independent (Z,A) regions as well as a simple linear optimization procedure whereby the predictive accuracy of the formula can be improved without altering the coefficients of the calibrated formula. Effectively this provides a possible means of ameliorating the statistical limitation of the narrower data set of the 'standard' without changing its characteristics. While this work has centred on the GLDM formula, the same methodology could be applied to other analytical formulae with coefficients derived from fits to experimental values.

*Acknowledgements*: The authors express their deep gratitude to Prof. Dr. Sudip Ghosh, Saha Institute of Nuclear Physics, Kolkata, India, for his critique. They also thank the University of Texas System Louis Stokes Alliance for Minority Participation in Research Programme for funding this work.

# EXPLANATION OF TABLES

**TABLE 1.** Atomic (Z) and mass (A) numbers, the experimental and calculated Q value of alpha decay ($Q_\alpha$), the experimental and calculated values of the alpha decay half-lives (T), and the standard error (SE) on the experimental alpha decay half-life, for the 77 REFERENCE nuclides.

This table lists the experimental data and calculated values for the 77 REFERENCE nuclides spanning the range of $^{146}_{62}Sm$ to $^{254}_{99}Es$. These nuclides are the alpha standards taken from [18] and a few others that also possess well-known alpha spectroscopic data. The four nucleon parity sets, e-e, e-o, o-e and o-o are labeled at the top each sub-section of the table. The experimental values were derived from [18] and the theoretical alpha half-life calculated from the Generalized Liquid Drop Model (GLDM), was taken from [14]. All Q values are in MeV and all half-lives are in seconds. The explanation of the entries in the table are given below.

Parent (column 1): The parent alpha decaying nuclide

Z,A (column 2): The charge and mass numbers of the parent nucleus.

$Q_\alpha$(exp) (column 3): The experimental Q value of alpha decay [18].

$Q_\alpha$(FRDM) (column 4): The Q value of alpha decay calculated using the Finite Range Droplet Model (FRDM) tabulation [20]

T(E) (column 5): The experimental partial half-life of alpha decay.

SE (T(E)) (column 6): The standard error on the experimental half-life of alpha decay. The values of T(E) and the propagated error SE(T(E)) were calculated from the experimental total half-life and $\alpha$ branching fraction and their standard deviations [18].

T(T-E) (column 7): The GLDM theory calculated $\alpha$ half-life [14].

RMFDsq(E) and RMFDsq(T) (bottom of sub-table for each nucleon parity set):

Root Mean Fractional Deviation square calculated as the square root of the mean of the fractional deviation square (eqn. 2) for all the nuclides in each parity set. E and T stand for the experimental and the theoretically calculated half-life.

**EXPLANATION OF TABLES** (*continued*)

**TABLE 2**.   Atomic and mass numbers**,** the experimental and calculated Q value of alpha decay, the experimental and calculated values of the alpha decay half-lives, and the standard error on the experimental alpha decay half-life, for the 225 TEST nuclides

This table lists the experimental data and calculated values for the 225 TEST nuclides spanning the range of $^{144}_{60}$Nd to $^{258}_{101}$Md. These nuclides cover approximately the same [Z, A] range as REFERENCE and possess well-known alpha spectroscopic data [18], but are not alpha standards. The four nucleon parity sets, e-e, e-o, o-e and o-o are labeled at the top each sub-section of the table. The experimental values were extracted from [18] and the calculated alpha half-lives were arrived at using the analytical GLDM formula [14], with coefficients as derived in this work in which both experimental and FRDM [20] $Q_\alpha$ values were employed. See the explanation of Table 1 for the expansion of the abbreviations. All Q values are in MeV and all half-lives are in seconds. The explanation of the entries in the table are given below.

Column 1 to Column 7:   Entries are the same as in Table 1

T(DSR-E) (column 8):   The DSR-E calculated α half-life. This variant of the GLDM analytical formula, possesses coefficients that have been derived using the $Q_\alpha$ (exp) of REFERENCE..
The calculated α half-life, T(DSR-E), of the unknown nuclide is obtained by using its $Q_\alpha$ (exp) in the resulting analytical formula.

$$\log_{10}[T(DSR\text{-}E)] = a + b.A^{1/6}.Z^{1/2} + c.Z/Q_\alpha(exp)^{1/2}$$

e-e nuclides

a = -22.267        b = - 1.1908        c = 1.5229

e-o nuclides

a = -27.7933        b = - 1.0031        c = 1.5813

o-e nuclides

a = -28.8157        b = - 1.0375        c = 1.6135

o-o nuclides

a = -28.2982        b = - 1.3555        c = 1.8265

T(DSR-FRDM) (column 9):
   The DSR-FRDM calculated α half–life (for e-e nuclides alone). Same as for T(DSR-E) except that it incorporates the FRDM $Q_\alpha$ in the calculation of the half-life in the DSR analytical formula.

$$\log_{10}[T(DSR\text{-}FRDM)] = a + b.A^{1/6}.Z^{1/2} + c.Z/Q_\alpha(FRDM)^{1/2}$$

where the values of a, b and c are the same as above.

T(FRDM-FRDM) (column 10):

The FRDM-FRDM calculated $\alpha$ half-life. This variant of the GLDM analytical formula, possesses coefficients that have been derived using the $Q_\alpha$ (FRDM) of REFERENCE. The calculated $\alpha$ half-life, T(FRDM-FRDM), of the unknown nuclide is obtained by using its $Q_\alpha$ (FRDM) in the resulting analytical formula.

$$\log_{10}[T(FRDM\text{-}FRDM)] = a + b.A^{1/6}.Z^{1/2} + c.Z/Q_\alpha(FRDM)^{1/2}$$

<u>e-e nuclides</u>

a = -7.2653     b = - 1.2130     c = 1.1441

<u>e-o nuclides</u>

a = 5.5997      b = -2.0451      c = 1.3673

<u>o-e nuclides</u>

a = -6.5247     b = - 1.5401     c = 1.3528

<u>o-o nuclides</u>

a = -29.0028    b = - 0.7217     c = 1.4534

RMFDsq(E), RMFDsq(T), RMFDsq(DSR-E), RMFDsq(DSR-FRDM), RMFDsq(FRDM-FRDM) (bottom of sub-table for each nucleon parity):

Same meaning as in table 1. The new terms in the parentheses stand for half-lives calculated using the DSR-E, the DSR-FRDM and the FRDM-FRDM formulae respectively.

**EXPLANATION OF TABLES**  (*continued*)

**TABLE 3**.   Atomic and mass numbers**,** the experimental and calculated Q value of alpha decay and the experimental and calculated values of the alpha decay half-lives, for the 22 heavy and super-heavy elements (SHE).

This table lists the experimental data and calculated values for the heavy and SHE nuclides spanning the range of $^{241}_{99}$Es to $^{294}_{118}$Uuo. The experimental values for the SHE were derived as follows: a) from [2], b) from [14] and the rest as before from [18]. The calculated α half-lives were arrived at using the analytical GLDM formula [14] with coefficients as derived in this work in which both experimental and FRDM [20] $Q_\alpha$ values were employed as well as the expressions resulting from the linear optimization of the calculated half-lives of TEST to its experimental values. See the explanation of Table 1 for the expansion of the abbreviations. All Q values are in MeV and all half-lives are in seconds. The explanation of the entries in the table, are given below.

Column 1 to Column 5:   Entries are the same as in Table 1

T(R-E ) (column 6):   The R-E calculated α half-life. This is Royer's [14] analytical formula derived from the GLDM. $Q_\alpha$(exp) values are used in the formula to calculate the unknown alpha half-lives.

T(DSR-E) (column 7):   Same as in table 2

T(FRDM-FRDM) (column 8):
    Same as in table 2

T(DSR-E-fit) (column 9):   The DSR-E-fit calculated α half-life. The half-lives of the TEST nuclides obtained from the DSR-E formula (see table 2) were linearly optimized to their experimental values. The fit equation and the T(DSR-E) for the heavy nuclides and SHE were used to obtain T(DSR-E-fit).

$$\log_{10}[T(DSR\text{-}E\text{-}fit)] \quad = \quad \mu \log_{10}[T(DSR\text{-}E)] \quad + \quad \chi$$

<u>e-e nuclides</u>

$\mu$   =   1.0306     $\chi$   =   - 0.3620

<u>e-o nuclides</u>

$\mu$   =   0.9911     $\chi$   =   - 0.2763

**EXPLANATION OF TABLES** (*continued*)

Table 3, T(DSR-E-fit) (column 9) (*continued*)

<u>o-e nuclides</u>

$\mu$ = 1.0431        $\chi$ = 0.3491

<u>o-o nuclides</u>

$\mu$ = 0.9093        $\chi$ = 0.5283

T(FRDM-FRDM-fit) (column 10):
> The FRDM-FRDM-fit calculated $\alpha$ half-life. Same as for the T(DSR-E-fit) for heavy nuclides and SHE except that the FRDM-FRDM formula is used.

$$\log_{10}[T(\text{FRDM-FRDM-fit})] = \mu \log_{10}[T(\text{FRDM-FRDM})] + \chi$$

<u>e-e nuclides</u>

$\mu$ = 1.1293        $\chi$ = $-2.3390$

<u>e-o nuclides</u>

$\mu$ = 0.8069        $\chi$ = $-2.6233$

<u>o-e nuclides</u>

$\mu$ = 0.9784        $\chi$ = $-2.0091$

<u>o-o nuclides</u>

$\mu$ = 1.0203        $\chi$ = $-1.0338$

RMFDsq(E), RMFDsq(T), RMFDsq(DSR-E), RMFDsq(FRDM-FRDM), RMSFDsq(DSR-E-fit) and RMFDsq(FRDM-FRDM-fit) (bottom of sub-table for each nucleon parity) :
> Same meaning as in table 1. The new terms in the parentheses stand for half-lives calculated using the DSR-E-fit and the FRDM-FRDM-fit formulae respectively.

**TABLE 1.** Atomic and mass numbers, the experimental and calculated Q value of alpha decay, the experimental and calculated values of the alpha decay half-lives, the standard error on the experimental alpha decay half-life, for the 77 REFERENCE nuclides.
See p. 22 for Explanation of Table.

| | | | Even-Even | | | |
|---|---|---|---|---|---|---|
| Parent | Z, A | $Q\alpha$ (exp) | $Q\alpha$ (FRDM) | T(E) | SE(T(E)) | T(T-E) |
| Sm | 62,146 | 2.53 | 3.15 | 3.25E15 | 1.58E14 | 7.24E15 |
| Gd | 64,148 | 3.27 | 3.83 | 2.35E09 | 9.47E07 | 4.79E09 |
| Dy | 66,154 | 2.95 | 2.29 | 9.47E13 | 4.73E13 | 1.17E14 |
| Po | 84,206 | 5.33 | 5.05 | 1.40E07 | 2.04E05 | 2.95E06 |
| Po | 84,208 | 5.22 | 4.79 | 9.15E07 | 6.31E04 | 1.17E07 |
| Po | 84,210 | 5.41 | 5.23 | 1.20E07 | 1.73E02 | - |
| Rn | 86,222 | 5.59 | 5.63 | 3.30E05 | 2.59E01 | 6.31E05 |
| Ra | 88,224 | 5.79 | 5.89 | 3.16E05 | 3.46E03 | 5.50E05 |
| Ra | 88,226 | 4.87 | 4.89 | 5.05E10 | 2.21E08 | 9.55E10 |
| Th | 90,228 | 5.52 | 5.53 | 6.04E07 | 2.84E04 | 1.17E08 |
| Th | 90,230 | 4.77 | 4.73 | 2.38E12 | 9.47E09 | 5.62E12 |
| Th | 90,232 | 4.08 | 3.90 | 4.34E17 | 1.89E15 | 1.51E18 |
| U | 92,230 | 5.99 | 5.65 | 1.80E06 | 0.00E00 | 3.47E06 |
| U | 92,232 | 5.41 | 5.15 | 2.17E09 | 1.26E07 | 4.90E09 |
| U | 92,234 | 4.86 | 4.71 | 7.75E12 | 1.89E10 | 1.62E13 |
| U | 92,236 | 4.57 | 4.48 | 7.39E14 | 9.47E11 | 1.70E15 |
| U | 92,238 | 4.27 | 4.26 | 1.41E17 | 9.47E13 | 4.27E17 |
| Pu | 94,236 | 5.87 | 5.78 | 9.02E07 | 2.52E05 | 1.02E08 |
| Pu | 94,238 | 5.59 | 5.59 | 2.77E09 | 9.47E06 | 3.47E09 |
| Pu | 94,240 | 5.26 | 5.18 | 2.07E11 | 2.21E08 | 3.24E11 |
| Pu | 94,242 | 4.98 | 4.80 | 1.18E13 | 3.79E10 | 2.19E13 |

**TABLE 1.** Atomic and mass numbers, the experimental and calculated Q value of alpha decay, the experimental and calculated values of the alpha decay half-lives, the standard error on the experimental alpha decay half-life, for the 77 REFERENCE nuclides.
See p. 22 for Explanation of Table.

Even-Even

| Parent | Z, A | Qα (exp) | Qα(FRDM) | T(E) | SE(T(E)) | T(T-E) |
|---|---|---|---|---|---|---|
| Pu | 94,244 | 4.66 | 4.55 | 2.55E15 | 3.16E13 | 3.55E15 |
| Cm | 96,242 | 6.22 | 6.17 | 1.41E07 | 1.73E04 | 1.29E07 |
| Cm | 96,244 | 5.90 | 5.66 | 5.71E08 | 6.31E05 | 4.79E08 |
| Cm | 96,248 | 5.16 | 4.92 | 1.17E13 | 1.38E11 | 1.20E13 |
| Cf | 98,246 | 6.86 | 6.50 | 1.29E05 | 1.80E03 | 8.32E04 |
| Cf | 98,248 | 6.36 | 6.27 | 2.88E07 | 2.42E05 | 1.70E07 |
| Cf | 98,250 | 6.13 | 5.88 | 4.13E08 | 2.84E06 | 2.34E08 |
| Cf | 98,252 | 6.22 | 6.26 | 8.61E07 | 2.61E05 | 7.59E07 |
| Cf | 98,254 | 5.93 | 6.00 | 1.69E09 | 1.09E08 | 2.14E09 |
| Fm | 100,252 | 7.15 | 6.80 | 9.14E04 | 1.80E02 | 3.02E04 |

RMFDsq(E) = 0.09;   RMFDsq(T) = 0.94;   Ratio of RMFDsq(T) to RMFDsq(E) is 10.28

**TABLE 1.** Atomic and mass numbers, the experimental and calculated Q value of alpha decay, the experimental and calculated values of the alpha decay half-lives, the standard error on the experimental alpha decay half-life, for the 77 REFERENCE nuclides.
See p. 22 for Explanation of Table

Even-Odd

| Parent | Z, A | $Q\alpha$ (exp) | $Q\alpha$(FRDM) | T(E) | SE(T(E)) | T(T-E) |
|---|---|---|---|---|---|---|
| Sm | 62,147 | 2.31 | 3.29 | 3.35E18 | 6.31E16 | 6.17E18 |
| Gd | 64,149 | 3.10 | 3.88 | 1.86E11 | 4.34E10 | 1.41E11 |
| Po | 84,209 | 4.98 | 4.70 | 3.23E09 | 1.59E08 | 2.51E08 |
| Ra | 88,223 | 5.98 | 6.12 | 9.88E05 | 3.46E02 | 2.24E05 |
| Th | 90,227 | 6.15 | 6.05 | 1.62E06 | 1.73E03 | - |
| Th | 90,229 | 5.17 | 5.18 | 2.32E11 | 5.05E09 | 4.57E11 |
| U | 92,233 | 4.91 | 4.91 | 5.02E12 | 6.31E09 | 7.41E12 |
| U | 92,235 | 4.68 | 4.56 | 2.22E16 | 1.58E13 | 8.71E15 |
| Pu | 94,239 | 5.25 | 5.32 | 7.61E11 | 9.47E08 | 4.47E11 |
| Pu | 94,241 | 5.14 | 5.02 | 1.85E13 | 1.98E11 | 2.24E13 |
| Cm | 96,241 | 6.19 | 6.31 | 2.83E08 | 2.84E07 | 1.10E08 |
| Cm | 96,243 | 6.17 | 5.89 | 9.21E08 | 3.18E06 | 7.41E08 |
| Cm | 96,245 | 5.62 | 5.42 | 2.68E11 | 3.16E09 | 1.78E11 |
| Cm | 96,247 | 5.35 | 5.08 | 4.92E14 | 1.58E13 | 3.47E14 |
| Cf | 98,249 | 6.30 | 6.11 | 1.11E10 | 6.31E07 | - |
| Cf | 98,251 | 6.18 | 6.15 | 2.83E10 | 1.39E09 | 2.09E10 |
| Cf | 98,253 | 6.12 | 6.45 | 4.96E08 | 6.41E07 | 2.09E08 |
| Fm | 100,253 | 7.20 | 7.18 | 2.16E06 | 2.00E05 | 6.92E04 |
| Fm | 100,255 | 7.24 | 7.19 | 7.23E04 | 2.52E02 | 3.02E04 |

RMFDsq(E) = 0.07;    RMFDsq(T) = 0.61;    Ratio of RMFDsq(T) to RMFDsq(E) is 8.53

**TABLE 1.** Atomic and mass numbers**,** the experimental and calculated Q value of alpha decay, the experimental and calculated values of the alpha decay half-lives, the standard error on the experimental alpha decay half-life, for the 77 REFERENCE nuclides.

Odd-Even

---

| Parent | Z, A | $Q\alpha$ (exp) | $Q\alpha$(FRDM) | T(E) | SE(T(E)) | T(T-E) |
|---|---|---|---|---|---|---|
| Eu | 63,147 | 2.99 | 3.50 | 9.46E10 | 2.59E10 | 3.63E11 |
| Ac | 89,225 | 5.94 | 6.23 | 8.64E05 | 8.64E03 | 3.31E05 |
| Ac | 89,227 | 5.04 | 5.18 | 4.98E10 | 1.44E08 | 2.63E10 |
| Pa | 91,229 | 5.84 | 5.86 | 2.70E07 | 2.95E06 | 4.90E07 |
| Pa | 91,231 | 5.15 | 5.08 | 1.03E12 | 3.47E09 | - |
| Np | 93,235 | 5.19 | 5.10 | 1.32E12 | 6.59E10 | 9.55E11 |
| Np | 93,237 | 4.96 | 4.90 | 6.75E13 | 3.16E11 | 3.89E13 |
| Am | 95,241 | 5.64 | 5.66 | 1.36E10 | 2.21E07 | 1.20E10 |
| Am | 95,243 | 5.44 | 5.41 | 2.33E11 | 1.26E09 | 2.40E11 |
| Bk | 97,245 | 6.46 | 6.09 | 3.56E08 | 2.97E07 | 4.68E08 |
| Bk | 97,249 | 5.53 | 5.29 | 1.91E12 | 1.11E11 | 2.34E11 |
| Es | 99,251 | 6.60 | 6.32 | 2.42E07 | 5.98E06 | 3.16E06 |
| Es | 99,253 | 6.74 | 6.76 | 1.77E06 | 2.59E03 | 5.75E05 |
| Es | 99,255 | 6.44 | 6.78 | 4.30E07 | 2.51E06 | 1.45E07 |

---

RMFDsq(E) = 0.11;    RMFDsq(T) = 0.96;    Ratio of RMFDsq(T) to RMFDsq(E) is 8.90

**TABLE 1.** Atomic and mass numbers**,** the experimental and calculated Q value of alpha decay, the experimental and calculated values of the alpha decay half-lives, the standard error on the experimental alpha decay half-life, for the 77 REFERENCE nuclides.

Odd-Odd

---

| Parent | Z, A | $Q\alpha$ (exp) | $Q\alpha$(FRDM) | T(E) | SE(T(E)) | T(T-E) |
|---|---|---|---|---|---|---|
| Ho | 67,152 | 4.51 | 4.50 | 1.34E03 | 3.71E02 | 7.24E02 |
| Tm | 69,156 | 4.34 | 4.06 | 1.31E05 | 2.07E04 | 7.94E04 |
| Ir | 77,172 | 5.99 | 5.88 | 2.20E02 | 1.50E01 | 3.16E00 |
| Au | 79,184 | 5.30 | 5.84 | 2.41E05 | 6.36E03 | 3.02E04 |
| Bi | 83,210 | 5.04 | 5.07 | 3.28E11 | 2.49E10 | - |
| At | 85,206 | 5.89 | 5.87 | 2.02E05 | 1.86E04 | 8.71E04 |
| At | 85,208 | 5.75 | 5.71 | 1.07E06 | 1.18E05 | 5.89E04 |
| Fr | 87,208 | 6.77 | 6.68 | 6.57E01 | 2.94E00 | 1.55E01 |
| Pa | 91,226 | 6.99 | 7.69 | 1.46E02 | 1.90E01 | 4.90E01 |
| Pa | 91,230 | 5.44 | 5.44 | 4.70E10 | 1.99E09 | - |
| Am | 95,240 | 5.69 | 5.80 | 9.63E10 | 3.55E10 | 1.10E10 |
| Es | 99,252 | 6.76 | 6.66 | 5.36E07 | 2.83E06 | 6.91E05 |
| Es | 99,254 | 6.62 | 6.95 | 2.38E07 | 4.32E04 | - |

---

RMFDsq(E) = 0.15;    RMFDsq(T) = 0.78;    Ratio of RMFDsq(T) to RMFDsq(E) is 5.34

**TABLE 2**. Atomic and mass numbers, the experimental and calculated Q value of alpha decay, the experimental and calculated values of the alpha decay half-lives, the standard error on the experimental alpha decay half-life, for the 225 TEST nuclides

Even-Even

| Parent | Z, A | $Q\alpha$ (exp) | $Q\alpha$(FRDM) | T(E) | SE(T(E)) | T(T-E) | T(DSR-E) | T(DSR-FRDM) | T(FRDM-FRDM) |
|---|---|---|---|---|---|---|---|---|---|
| Nd | 60,144 | 1.91 | 2.44 | 7.23E22 | 5.05E21 | 4.17E23 | 6.57E22 | 1.49E15 | 1.64E15 |
| Sm | 62,148 | 1.99 | 2.46 | 2.21E23 | 9.47E22 | 7.94E23 | 1.49E23 | 2.69E16 | 1.09E16 |
| Gd | 64,150 | 2.81 | 3.05 | 5.65E13 | 2.53E12 | 1.23E14 | 8.49E13 | 4.25E11 | 2.13E12 |
| Gd | 64,152 | 2.21 | 1.42 | 3.41E21 | 2.53E20 | 1.32E22 | 2.31E21 | 4.57E37 | 7.42E31 |
| Dy | 66,150 | 4.35 | 4.40 | 1.20E03 | 1.66E02 | 1.66E03 | 4.17E03 | 2.39E03 | 1.09E06 |
| Dy | 66,152 | 3.73 | 3.72 | 8.57E06 | 6.04E05 | 1.78E07 | 2.81E07 | 3.41E07 | 1.40E09 |
| Er | 68,152 | 4.93 | 5.22 | 1.14E01 | 5.21E-01 | 1.62E01 | 4.66E01 | 2.49E00 | 4.95E03 |
| Er | 68,154 | 4.28 | 4.55 | 4.76E04 | 1.32E04 | 5.75E04 | 1.14E05 | 3.75E03 | 1.17E06 |
| Yb | 70,154 | 5.47 | 5.72 | 4.35E-01 | 1.78E-02 | 5.13E-01 | 1.70E00 | 1.82E-01 | 5.49E02 |
| Yb | 70,156 | 4.81 | 5.20 | 2.61E02 | 5.27E01 | 6.31E02 | 1.64E03 | 2.45E01 | 2.12E04 |
| Hf | 72,158 | 5.40 | 5.91 | 6.50E00 | 5.73E-01 | 8.51E00 | 2.59E01 | 2.31E-01 | 5.04E02 |
| Hf | 72,160 | 4.90 | 4.89 | 5.65E02 | 1.61E02 | - | 5.13E03 | 6.33E03 | 1.06E06 |

**TABLE 2**. Atomic and mass numbers, the experimental and calculated Q value of alpha decay, the experimental and calculated values of the alpha decay half-lives, the standard error on the experimental alpha decay half-life, for the 225 TEST nuclides

Even-Even

| Parent | Z, A | $Q\alpha$ (exp) | $Q\alpha$(FRDM) | T(E) | SE(T(E)) | T(T-E) | T(DSR-E) | T(DSR-FRDM) | T(FRDM-FRDM) |
|---|---|---|---|---|---|---|---|---|---|
| Hf | 72,162 | 4.42 | 4.41 | 5.97E05 | 1.33E05 | 1.20E06 | 2.07E06 | 2.43E06 | 8.95E07 |
| Hf | 72,174 | 2.50 | 2.40 | 6.31E22 | 1.26E22 | - | 1.84E23 | 5.13E24 | 4.40E21 |
| W | 74,162 | 5.68 | 5.53 | 2.96E00 | 2.07E-01 | 4.57E00 | 1.34E01 | 5.77E01 | 2.46E04 |
| W | 74,164 | 5.28 | 5.21 | 2.31E02 | 1.51E02 | 2.40E02 | 6.57E02 | 1.46E03 | 2.70E05 |
| W | 74,168 | 4.51 | 4.38 | 1.89E06 | 3.58E05 | - | 5.77E06 | 3.54E07 | 5.02E08 |
| Os | 76,166 | 6.13 | 5.99 | 2.51E-01 | 6.96E-02 | 3.98E-01 | 1.38E00 | 5.09E00 | 3.07E03 |
| Os | 76,168 | 5.82 | 5.67 | 4.49E00 | 3.42E-01 | 7.08E00 | 2.13E01 | 9.47E01 | 2.67E04 |
| Os | 76,170 | 5.54 | 5.34 | 5.92E01 | 5.21E00 | 1.05E02 | 2.98E02 | 2.57E03 | 3.10E05 |
| Os, | 76,172 | 5.23 | 4.96 | 9.50E03 | 1.00E03 | 3.16E03 | 7.54E03 | 1.77E05 | 7.23E06 |
| Os | 76,186 | 2.82 | 2.86 | 6.31E22 | 3.47E22 | 4.57E22 | 6.76E21 | 2.69E21 | 8.51E18 |
| Pt | 78,174 | 6.18 | 6.31 | 1.08E00 | 6.64E-02 | 1.45E00 | 4.49E00 | 1.55E00 | 9.17E02 |
| Pt | 78,176 | 5.89 | 6.03 | 1.67E01 | 1.37E00 | 2.19E01 | 6.30E01 | 1.70E01 | 5.38E03 |
| Pt | 78,190 | 3.25 | 4.56 | 2.05E19 | 9.47E17 | 1.02E19 | 2.63E18 | 1.13E08 | 5.89E08 |

**TABLE 2**. Atomic and mass numbers**,** the experimental and calculated Q value of alpha decay, the experimental and calculated values of the alpha decay half-lives, the standard error on the experimental alpha decay half-life, for the 225 TEST nuclides
See p. 23 for Explanation of Table.

Even-Even

| Parent | Z, A | Qα (exp) | Qα(FRDM) | T(E) | SE(T(E)) | T(T-E) | T(DSR-E) | T(DSR-FRDM) | T(FRDM-FRDM) |
|---|---|---|---|---|---|---|---|---|---|
| Hg | 80,180 | 6.26 | 6.64 | 5.83E00 | 6.42E-01 | - | 1.34E01 | 5.29E-01 | 3.08E02 |
| Hg | 80,182 | 6.00 | 6.35 | 7.13E01 | 3.76E00 | 4.79E01 | 1.33E02 | 5.57E00 | 1.75E03 |
| Hg | 80,184 | 5.66 | 5.83 | 2.76E03 | 1.51E02 | 1.58E03 | 3.42E03 | 6.49E02 | 6.07E04 |
| Hg | 80,186 | 5.21 | 5.47 | 5.16E05 | 1.64E05 | 2.51E05 | 4.82E05 | 2.53E04 | 9.24E05 |
| Hg | 80,188 | 4.71 | 5.09 | 5.27E08 | 1.17E08 | - | 2.39E08 | 1.86E06 | 2.27E07 |
| Pb | 82,188 | 6.11 | 5.73 | 1.10E02 | 3.53E01 | - | 2.76E02 | 1.31E04 | 4.51E05 |
| Pb | 82,190 | 5.70 | 5.29 | 8.00E03 | 1.90E03 | - | 1.56E04 | 1.58E06 | 1.61E07 |
| Pb | 82,192 | 5.22 | 4.76 | 3.68E06 | 6.55E05 | - | 3.07E06 | 1.26E09 | 2.37E09 |
| Po | 84,198 | 6.31 | 6.84 | 1.85E02 | 7.23E00 | 8.13E01 | 2.07E02 | 2.07E00 | 4.52E02 |
| Po | 84,200 | 5.98 | 6.37 | 6.22E03 | 1.77E02 | 1.91E03 | 4.44E03 | 1.11E02 | 8.77E03 |
| Po | 84,202 | 5.70 | 5.97 | 1.34E05 | 1.35E04 | 3.89E04 | 7.49E04 | 4.73E03 | 1.43E05 |
| Po | 84,204 | 5.48 | 5.55 | 1.93E06 | 3.12E04 | 4.90E05 | 7.52E05 | 3.79E05 | 3.75E06 |
| Po | 84,212 | 8.95 | 7.79 | 2.99E-07 | 2.00E-09 | 1.45E-07 | 6.83E07 | 8.55E-04 | 1.08E00 |

**TABLE 2**. Atomic and mass numbers, the experimental and calculated Q value of alpha decay, the experimental and calculated values of the alpha decay half-lives, the standard error on the experimental alpha decay half-life, for the 225 TEST nuclides
See p. 23 for Explanation of Table.

Even-Even

| Parent | Z, A | $Q\alpha$ (exp) | $Q\alpha$(FRDM) | T(E) | SE(T(E)) | T(T-E) | T(DSR-E) | T(DSR-FRDM) | T(FRDM-FRDM) |
|---|---|---|---|---|---|---|---|---|---|
| Po | 84,214 | 7.83 | 7.92 | 1.64E-04 | 2.00E-06 | 1.51E-04 | 5.60E-04 | 3.25E-04 | 5.07E-01 |
| Po | 84,216 | 6.91 | 7.26 | 1.45E-01 | 2.00E-03 | 1.82E-01 | 4.76E-01 | 3.12E-02 | 1.52E01 |
| Po | 84,218 | 6.11 | 6.05 | 1.86E02 | 6.00E-01 | 2.34E02 | 4.93E02 | 9.76E02 | 3.54E04 |
| Rn | 86,202 | 6.77 | 7.01 | 1.16E01 | 2.06E00 | 6.92E00 | 2.03E01 | 2.94E00 | 4.60E02 |
| Rn | 86,204 | 6.55 | 6.69 | 1.02E02 | 2.83E00 | 4.90E01 | 1.35E02 | 3.94E01 | 3.14E03 |
| Rn | 86,206 | 6.38 | 6.52 | 5.49E02 | 3.12E01 | 2.34E02 | 5.41E02 | 1.62E02 | 8.83E03 |
| Rn | 86,208 | 6.26 | 6.26 | 2.36E03 | 2.66E02 | 7.24E02 | 1.58E03 | 1.66E03 | 4.95E04 |
| Rn | 86,210 | 6.16 | 6.14 | 9.00E03 | 3.87E02 | 1.70E03 | 3.86E03 | 4.87E03 | 1.08E05 |
| Rn | 86,212 | 6.39 | 6.55 | 1.43E03 | 7.20E01 | 1.78E02 | 3.98E02 | 9.16E01 | 5.33E03 |
| Rn | 86,214 | 9.21 | 9.14 | 2.70E-07 | 2.00E-08 | 1.51E-07 | 7.71E-07 | 1.15E-06 | 6.00E-03 |
| Rn | 86,216 | 8.20 | 9.22 | 4.50E-05 | 5.00E-06 | 6.17E-05 | 2.64E-04 | 6.74E-07 | 3.92E-03 |
| Rn | 86,218 | 7.26 | 7.96 | 3.50E-02 | 5.00E-03 | 5.50E-02 | 1.74E-01 | 1.20E-03 | 1.06E00 |
| Rn | 86,220 | 6.40 | 6.36 | 5.56E01 | 1.00E-01 | 1.05E02 | 2.25E02 | 3.59E02 | 1.34E04 |

**TABLE 2**. Atomic and mass numbers, the experimental and calculated Q value of alpha decay, the experimental and calculated values of the alpha decay half-lives, the standard error on the experimental alpha decay half-life, for the 225 TEST nuclides

Even-Even

| Parent | Z, A | Qα (exp) | Qα(FRDM) | T(E) | SE(T(E)) | T(T-E) | T(DSR-E) | T(DSR-FRDM) | T(FRDM-FRDM) |
|---|---|---|---|---|---|---|---|---|---|
| Ra | 88,208 | 7.27 | 7.02 | 1.37E00 | 2.11E-01 | 5.50E-01 | 1.72E00 | 1.39E01 | 1.12E03 |
| Ra | 88,214 | 7.27 | 7.23 | 2.46E00 | 3.00E-02 | 4.17E-01 | 1.28E00 | 1.87E00 | 2.30E02 |
| Ra | 88,216 | 9.53 | 9.69 | 1.82E-07 | 1.00E-08 | 1.17E-07 | 6.21E-07 | 2.72E-07 | 1.64E-03 |
| Ra | 88,218 | 8.55 | 9.71 | 2.56E-05 | 1.100E-06 | 3.24E-05 | 1.49E-04 | 2.23E-07 | 1.37E-03 |
| Ra | 88,220 | 7.60 | 7.36 | 2.5E-02 | 5.00E-03 | 2.34E-02 | 8.25E-02 | 5.05E-01 | 7.96E01 |
| Ra | 88,222 | 6.68 | 6.00 | 3.80E01 | 5.00E-01 | 5.25E01 | 1.24E02 | 9.56E04 | 7.15E05 |
| Th | 90,212 | 7.95 | 7.96 | 3.00E-02 | 1.50E-02 | - | 5.67E-02 | 5.54E-02 | 1.39E01 |
| Th | 90,214 | 7.83 | 7.55 | 1.00E-01 | 2.5E-02 | 3.31E-02 | 1.25E-01 | 1.01E00 | 1.20E02 |
| Th | 90,216 | 8.07 | 7.81 | 2.80E-02 | 2.00E-03 | 5.25E-03 | 2.03E-02 | 1.33E-01 | 2.54E01 |
| Th | 90,218 | 9.85 | 10.16 | 1.09E-07 | 1.3E-08 | 9.12E-08 | 4.94E-07 | 1.07E-07 | 6.55E-04 |
| Th | 90,220 | 8.95 | 9.83 | 9.7E-06 | 6.00E-07 | 1.29E-05 | 6.09E-05 | 5.06E-07 | 2.05E-03 |
| Th | 90,222 | 8.13 | 7.07 | 2.80E-03 | 3.00E-04 | 2.63E-03 | 1.02E-02 | 3.17E01 | 1.44E03 |
| Th | 90,224 | 7.30 | 7.30 | 1.05E00 | 2.00E-02 | 1.48E00 | 4.07E00 | 4.37E00 | 3.16E02 |

**TABLE 2**. Atomic and mass numbers, the experimental and calculated Q value of alpha decay, the experimental and calculated values of the alpha decay half-lives, the standard error on the experimental alpha decay half-life, for the 225 TEST nuclides
See p. 23 for Explanation of Table

Even-Even

| Parent | Z, A | $Q\alpha$ (exp) | $Q\alpha$(FRDM) | T(E) | SE(T(E)) | T(T-E) | T(DSR-E) | T(DSR-FRDM) | T(FRDM-FRDM) |
|---|---|---|---|---|---|---|---|---|---|
| Th | 90,226 | 6.45 | 7.02 | 1.85E03 | 0.00E00 | 3.09E03 | 6.53E03 | 4.00E01 | 1.63E03 |
| U  | 92,226 | 7.71 | 8.18 | 5.00E-01 | 2.00E-01 | - | 1.03E00 | 3.52E-02 | 6.82E00 |
| Cm | 96,246 | 5.47 | 5.18 | 1.49E11 | 3.17E09 | 1.32E11 | 1.03E11 | 6.24E12 | 1.85E11 |
| Cf | 98,244 | 7.33 | 6.91 | 1.16E03 | 3.60E01 | 9.77E02 | 2.48E03 | 1.15E05 | 2.43E05 |
| Fm | 100,246 | 8.37 | 7.83 | 1.20E00 | 2.21E-01 | 1.07E00 | 3.58E00 | 2.33E02 | 1.87E03 |
| Fm | 100,248 | 8.00 | 7.64 | 3.64E01 | 3.05E00 | 1.86E01 | 5.28E01 | 1.00E03 | 5.45E03 |
| Fm | 100,254 | 7.31 | 7.40 | 1.17E04 | 7.20E00 | 5.62E03 | 1.27E04 | 5.87E03 | 1.91E04 |
| Fm | 100,256 | 7.03 | 7.24 | 1.17E05 | 4.43E03 | 7.76E04 | 1.51E05 | 2.22E04 | 5.06E04 |
| No | 102,252 | 8.55 | 8.35 | 3.15E00 | 3.12E-01 | 1.29E00 | 4.33E00 | 1.91E01 | 2.20E02 |
| No | 102,254 | 8.23 | 7.97 | 6.11E01 | 4.30E00 | 1.32E01 | 4.26E01 | 3.23E02 | 1.80E03 |
| No | 102,256 | 8.58 | 8.57 | 3.32E00 | 2.01E-01 | 8.51E-01 | 2.87E00 | 3.21E00 | 5.50E01 |

RMFDsq(E) = 0.18; RMFDsq(T) = 0.94; RMFDsq(DSR-E) = 2.35; RMFDsq(DSR-FRDM) = 1.54E15; RMFDsq(FRDM-FRDM) = 2.50E09

**TABLE 2**. Atomic and mass numbers, the experimental and calculated Q value of alpha decay, the experimental and calculated values of the alpha decay half-lives, the standard error on the experimental alpha decay half-life, for the 225 TEST nuclides
See p. 23 for Explanation of Table

Even-Odd

| Parent | Z, A | Qα (exp) | Qα(FRDM) | T(E) | SE(T(E)) | T(T-E) | T(DSR-E) | T(FRDM-FRDM) |
|---|---|---|---|---|---|---|---|---|
| Dy | 66,153 | 3.56 | 2.80 | 2.45E08 | 3.67E07 | 3.39E08 | 4.83E08 | 1.43E21 |
| Er | 68,153 | 4.80 | 5.06 | 7.00E01 | 3.98E00 | 7.08E01 | 1.40E02 | 8.97E07 |
| Er | 68,155 | 4.12 | 3.75 | 1.45E06 | 4.67E05 | 5.75E05 | 1.03E06 | 3.63E14 |
| Yb | 70,155 | 5.34 | 5.74 | 1.97E00 | 1.05E-01 | 1.86E00 | 4.70E00 | 8.14E05 |
| Yb | 70,157 | 4.62 | 4.61 | 7.72E03 | 2.00E02 | 6.61E03 | 1.64E04 | 2.89E10 |
| Hf | 72,157 | 5.88 | 6.06 | 1.28E-01 | 1.51E-02 | 7.94E-02 | 2.47E-01 | 2.01E05 |
| Hf | 72,159 | 5.22 | 5.01 | 1.37E01 | 1.93E00 | 5.62E01 | 1.70E02 | 1.64E09 |
| Hf | 72,161 | 4.72 | 4.71 | 5.79E03 | 1.04E03 | 2.40E04 | 5.78E04 | 3.25E10 |
| W | 74,161 | 5.92 | 5.97 | 5.00E-01 | 1.66E-01 | - | 1.50E00 | 9.90E05 |
| W | 74,163 | 5.52 | 5.41 | 6.71E00 | 1.02E00 | 1.86E01 | 7.03E01 | 1.01E08 |
| Os | 76,167 | 5.98 | 5.81 | 1.24E00 | 2.44E-01 | 1.55E00 | 6.80E00 | 7.81E06 |
| Os | 76,169 | 5.72 | 5.56 | 3.09E01 | 3.35E00 | 1.86E01 | 7.88E01 | 5.88E07 |
| Os | 76,171 | 5.37 | 5.12 | 4.71E02 | 9.27E01 | 6.46E02 | 2.94E03 | 3.50E09 |

**TABLE 2**. Atomic and mass numbers, the experimental and calculated Q value of alpha decay, the experimental and calculated values of the alpha decay half-lives, the standard error on the experimental alpha decay half-life, for the 225 TEST nuclides

Even-Odd

| Parent | Z, A   | $Q\alpha$ (exp) | $Q\alpha$(FRDM) | T(E)    | SE(T(E)) | T(T-E)  | T(DSR-E) | T(FRDM-FRDM) |
|--------|--------|-----------------|-----------------|---------|----------|---------|----------|--------------|
| Os     | 76,173 | 5.06            | 4.77            | 7.62E04 | 4.04E04  | 2.34E04 | 9.84E04  | 1.32E11      |
| Pt     | 78,173 | 6.35            | 6.37            | 4.07E-01| 3.61E-02 | 3.16E-01| 1.75E00  | 1.73E05      |
| Pt     | 78,175 | 6.18            | 6.10            | 3.94E00 | 3.32E-01 | 1.38E00 | 7.52E00  | 1.21E06      |
| Pt     | 78,177 | 5.64            | 5.92            | 1.96E02 | 2.27E01  | 2.51E02 | 1.36E03  | 4.51E06      |
| Pt     | 78,179 | 5.40            | 5.52            | 8.83E03 | 1.12E03  | 1.35E04 | 1.91E04  | 1.37E08      |
| Pt     | 78,185 | 4.54            | 4.69            | 8.51E07 | 3.42E07  | -       | 8.90E08  | 5.74E11      |
| Hg     | 80,175 | 7.04            | 7.00            | 2.00E-02| 2.70E-02 | -       | 4.66E-02 | 4.99E03      |
| Hg     | 80,177 | 6.74            | 6.97            | 1.53E-01| 5.88E-03 | 6.76E-02| 4.76E-01 | 5.07E03      |
| Hg     | 80,181 | 6.29            | 6.48            | 1.00E01 | 1.39E00  | 3.09E00 | 2.10E01  | 1.21E05      |
| Hg     | 80,183 | 6.04            | 6.13            | 8.03E01 | 1.50E01  | 3.16E01 | 2.04E02  | 1.64E06      |
| Hg     | 80,185 | 5.78            | 5.70            | 8.18E02 | 1.37E02  | -       | 2.63E03  | 5.94E07      |
| Pb     | 82,187 | 6.40            | 5.92            | 9.15E02 | 1.50E01  | -       | 5.80E01  | 2.60E07      |
| Po     | 84,195 | 6.75            | 7.27            | 6.19E00 | 1.24E00  | 1.51E00 | 1.57E01  | 1.19E03      |

**TABLE 2**. Atomic and mass numbers, the experimental and calculated Q value of alpha decay, the experimental and calculated values of the alpha decay half-lives, the standard error on the experimental alpha decay half-life, for the 225 TEST nuclides
See p. 23 for Explanation of Table

Even-Odd

| Parent | Z, A | $Q\alpha$ (exp) | $Q\alpha$(FRDM) | T(E) | SE(T(E)) | T(T-E) | T(DSR-E) | T(FRDM-FRDM) |
|---|---|---|---|---|---|---|---|---|
| Po | 84,197 | 6.41 | 6.87 | 1.27E02 | 2.14E01 | 3.16E01 | 3.14E02 | 1.67E04 |
| Po | 84,199 | 6.07 | 6.56 | 2.74E03 | 4.64E02 | 8.71E02 | 7.77E03 | 1.49E05 |
| Po | 84,201 | 5.80 | 6.14 | 5.74E04 | 1.08E04 | 1.29E04 | 1.31E05 | 4.04E06 |
| Po | 84,203 | 5.50 | 5.68 | 2.00E06 | 3.65E05 | 3.89E05 | 3.81E06 | 2.37E08 |
| Po | 84,205 | 5.32 | 5.33 | 1.49E07 | 3.74E06 | 3.47E06 | 2.83E07 | 7.23E09 |
| Po | 84,207 | 5.22 | 4.85 | 9.94E07 | 9.48E06 | 1.20E07 | 1.02E08 | 1.56E12 |
| Po | 84,211 | 7.59 | 6.76 | 5.16E-01 | 3.00E-03 | - | 9.49E-03 | 1.14E04 |
| Po | 84,213 | 8.54 | 7.98 | 4.20E-06 | 8.00E-07 | 1.55E-06 | 1.60E-05 | 2.93E00 |
| Po | 84,215 | 7.53 | 7.65 | 1.78E-03 | 4.00E-06 | 1.29E-03 | 1.33E-02 | 1.83E01 |
| Rn | 86,203 | 6.63 | 6.76 | 6.82E01 | 1.03E01 | 2.34E01 | 2.97E02 | 7.36E04 |
| Rn | 86,205 | 6.39 | 6.36 | 7.30E02 | 1.30E02 | 2.19E02 | 2.62E03 | 1.56E06 |
| Rn | 86,207 | 6.25 | 6.22 | 2.64E03 | 3.81E02 | 8.51E02 | 9.48E03 | 4.38E06 |
| Rn | 86,209 | 6.16 | 6.14 | 1.01E04 | 1.23E03 | 1.91E03 | 2.30E04 | 7.48E06 |

**TABLE 2**. Atomic and mass numbers, the experimental and calculated Q value of alpha decay, the experimental and calculated values of the alpha decay half-lives, the standard error on the experimental alpha decay half-life, for the 225 TEST nuclides

See p. 23 for Explanation of Table

Even-Odd

---

| Parent | Z, A | Qα (exp) | Qα(FRDM) | T(E) | SE(T(E)) | T(T-E) | T(DSR-E) | T(FRDM-FRDM) |
|---|---|---|---|---|---|---|---|---|
| Rn | 86,211 | 5.97 | 6.02 | 1.92E05 | 1.22E04 | 2.95E04 | 1.56E05 | 1.87E07 |
| Rn | 86,213 | 8.24 | 8.19 | 2.50E-02 | 2.00E-04 | - | 6.96E-04 | 2.26E00 |
| Rn | 86,215 | 8.84 | 9.38 | 2.30E-06 | 1.00E-07 | 1.20E-06 | 1.52E-05 | 3.85E-03 |
| Rn | 86,217 | 7.89 | 8.49 | 5.40E-04 | 5.00E-05 | 5.01E-04 | 6.66E-03 | 3.00E-01 |
| Rn | 86,219 | 6.95 | 6.87 | 3.96E00 | 1.00E-02 | 6.92E-01 | 9.34E00 | 8.27E03 |
| Rn | 86,221 | 6.15 | 5.79 | 6.82E03 | 6.27E02 | 1.26E03 | 1.55E04 | 7.20E07 |
| Ra | 88,213 | 6.86 | 6.71 | 2.06E02 | 1.36E01 | 3.98E01 | 2.22E02 | 1.53E05 |
| Ra | 88,215 | 8.86 | 8.79 | 1.59E-03 | 9.00E-05 | - | 8.26E-05 | 1.74E-01 |
| Ra | 88,217 | 9.16 | 9.80 | 1.60E-06 | 2.00E-07 | 8.32E-07 | 1.31E-05 | 1.04E-03 |
| Ra | 88,219 | 8.13 | 7.98 | 1.00E-02 | 3.00E-03 | - | 8.14E-03 | 1.28E01 |
| Ra | 88,221 | 6.89 | 6.57 | 2.80E01 | 2.00E00 | 3.39E01 | 1.26E02 | 2.45E05 |
| Th | 90,215 | 7.67 | 7.35 | 1.20E00 | 2.00E-01 | 3.02E-01 | 2.08E00 | 3.31E03 |
| Th | 90,217 | 9.42 | 9.27 | 2.52E-04 | 7.00E-06 | - | 1.74E-05 | 2.95E-02 |
| Th | 90,219 | 9.51 | 10.05 | 1.05E-06 | 3.00E-08 | 5.13E-07 | 9.89E-06 | 6.24E-04 |

**TABLE 2**. Atomic and mass numbers, the experimental and calculated Q value of alpha decay, the experimental and calculated values of the alpha decay half-lives, the standard error on the experimental alpha decay half-life, for the 225 TEST nuclides

See p. 23 for Explanation of Table

Even-Odd

| Parent | Z, A | $Q\alpha$ (exp) | $Q\alpha$(FRDM) | T(E) | SE(T(E)) | T(T-E) | T(DSR-E) | T(FRDM-FRDM) |
|---|---|---|---|---|---|---|---|---|
| Th | 90,221 | 8.63 | 7.90 | 1.68E-03 | 6.00E-05 | 8.32E-04 | 1.83E-03 | 4.92E01 |
| Th | 90,223 | 7.57 | 7.23 | 6.00E-01 | 2.00E-02 | 5.62E-01 | 3.25E00 | 4.03E03 |
| U | 92,225 | 8.02 | 8.35 | 9.50E-02 | 1.50E-02 | - | 7.11E-01 | 5.88E00 |
| U | 92,227 | 7.21 | 7.65 | 6.60E01 | 6.00E00 | 8.91E02 | 4.19E02 | 4.45E02 |
| Pu | 94,233 | 6.42 | 6.49 | 1.05E06 | 4.36E05 | 2.40E05 | 5.61E06 | 7.70E06 |
| Pu | 94,237 | 5.75 | 5.69 | 9.30E10 | 8.86E09 | 2.34E10 | 1.01E10 | 1.52E10 |
| Cf | 98,245 | 7.26 | 6.71 | 7.50E03 | 6.73E02 | 1.82E03 | 7.96E04 | 5.08E06 |
| Cf | 98,247 | 6.53 | 6.45 | 3.20E07 | 4.58E06 | - | 9.85E07 | 4.69E07 |
| Fm | 100,251 | 7.43 | 6.88 | 1.06E06 | 7.82E04 | 2.51E05 | 1.12E05 | 2.42E06 |
| Fm | 100,257 | 6.86 | 6.59 | 8.70E06 | 1.73E04 | 5.37E06 | 1.88E07 | 2.08E07 |
| No | 102,255 | 8.45 | 8.57 | 3.03E02 | 2.31E01 | 1.10E01 | 1.59E02 | 1.75E01 |

RMFDsq(E) = 0.23;  RMFDsq(T) = 1.91;  RMFDsq(DSR-E) = 4.74 ;  RMFDsq(FRDM-FRDM) = 7.18E11

**TABLE 2**. Atomic and mass numbers**,** the experimental and calculated Q value of alpha decay, the experimental and calculated values of the alpha decay half-lives, the standard error on the experimental alpha decay half-life, for the 225 TEST nuclides

Odd-Even

| Parent | Z, A | Qα (exp) | Qα(FRDM) | T(E) | SE(T(E)) | T(T-E) | T(DSR-E) | T(FRDM-FRDM) |
|---|---|---|---|---|---|---|---|---|
| Tb | 65,149 | 4.08 | 4.18 | 8.88E04 | 9.05E03 | 1.86E04 | 7.32E03 | 8.37E07 |
| Tb | 65,151 | 3.50 | 3.23 | 6.67E08 | 1.05E08 | 2.29E08 | 9.42E07 | 6.15E13 |
| Tm | 69,153 | 5.25 | 5.44 | 1.63E00 | 5.47E-02 | 1.58E00 | 7.11E-01 | 8.52E03 |
| Tm | 69,155 | 4.57 | 4.72 | 1.14E03 | 1.80E02 | 4.17E03 | 1.97E03 | 6.53E06 |
| Ta | 73,159 | 5.75 | 6.00 | 7.13E-01 | 2.29E-01 | 7.59E-01 | 4.70E-01 | 1.53E03 |
| Re | 75,163 | 6.07 | 5.87 | 4.06E-01 | 1.30E-01 | - | 2.01E-01 | 1.58E04 |
| Re | 75,165 | 5.66 | 5.52 | 1.85E01 | 6.28E00 | - | 1.02E01 | 2.79E05 |
| Ir | 77,177 | 5.13 | 5.27 | 5.00E04 | 8.98E03 | - | 2.92E04 | 7.11E06 |
| Au | 79,175 | 6.78 | 6.94 | 2.13E-01 | 4.31E-02 | - | 2.17E-02 | 4.85E01 |
| Au | 79,179 | 6.08 | 6.42 | 3.23E01 | 1.90E00 | 8.13E00 | 9.53E00 | 1.50E03 |
| Au | 79,181 | 5.75 | 6.10 | 8.77E02 | 1.40E02 | 2.14E02 | 2.52E02 | 1.62E04 |

| Parent | Z, A | Qα (exp) | Qα(FRDM) | T(E) | SE(T(E)) | T(T-E) | T(DSR-E) | T(FRDM-FRDM) |
|---|---|---|---|---|---|---|---|---|
| Au | 79,183 | 5.47 | 6.03 | 1.40E04 | 2.37E03 | 4.27E03 | 5.17E03 | 2.52E04 |

**TABLE 2**. Atomic and mass numbers**,** the experimental and calculated Q value of alpha decay, the experimental and calculated values of the alpha decay half-lives, the standard error on the experimental alpha decay half-life, for the 225 TEST nuclides

Odd-Even

| Parent | Z, A | Qα (exp) | Qα(FRDM) | T(E) | SE(T(E)) | T(T-E) | T(DSR-E) | T(FRDM-FRDM) |
|---|---|---|---|---|---|---|---|---|
| Au | 79,185 | 5.18 | 5.74 | 9.81E04 | 2.27E04 | 1.10E05 | 1.51E05 | 2.68E05 |
| Bi | 83,191 | 6.78 | 6.78 | 2.00E01 | 6.87E00 | 9.55E00 | 8.51E-01 | 8.80E02 |
| Bi | 83,193 | 6.31 | 6.29 | 1.34E03 | 8.06E02 | 5.01E02 | 6.26E01 | 3.43E04 |
| Bi | 83,195 | 5.83 | 5.82 | 5.55E05 | 3.70E05 | 8.71E04 | 7.47E03 | 1.79E06 |
| Bi | 83,211 | 6.75 | 6.24 | 1.29E02 | 1.20E00 | - | 4.64E-01 | 1.62E04 |
| Bi | 83,213 | 5.98 | 6.29 | 1.31E05 | 1.89E03 | - | 6.94E02 | 9.48E03 |
| At | 85,197 | 7.10 | 7.26 | 3.65E-01 | 4.43E-02 | 1.95E-01 | 3.82E-01 | 8.28E01 |
| At | 85,199 | 6.78 | 7.19 | 8.00E00 | 7.11E-01 | 2.82E00 | 5.55E00 | 1.17E02 |
| At | 85,201 | 6.47 | 6.90 | 1.25E02 | 4.23E00 | 3.98E01 | 8.68E01 | 8.01E02 |
| At | 85,203 | 6.21 | 6.63 | 1.43E03 | 1.44E02 | 4.79E02 | 1.07E03 | 5.37E03 |
| At | 85,205 | 6.02 | 6.35 | 1.57E04 | 3.16E03 | 3.24E03 | 7.19E03 | 4.46E04 |

| Parent | Z, A | Qα (exp) | Qα(FRDM) | T(E) | SE(T(E)) | T(T-E) | T(DSR-E) | T(FRDM-FRDM) |
|---|---|---|---|---|---|---|---|---|
| At | 85,207 | 5.87 | 5.81 | 7.53E04 | 8.92E03 | 1.58E04 | 3.26E04 | 4.67E06 |
| At | 85,209 | 5.76 | 5.61 | 4.75E05 | 5.81E04 | 4.37E04 | 1.10E05 | 2.87E07 |

**TABLE 2**. Atomic and mass numbers**,** the experimental and calculated Q value of alpha decay, the experimental and calculated values of the alpha decay half-lives, the standard error on the experimental alpha decay half-life, for the 225 TEST nuclides

Odd-Even

--------------------------------------------------------------------------------------------------------------------------------------------------

| Parent | Z, A | Qα (exp) | Qα(FRDM) | T(E) | SE(T(E)) | T(T-E) | T(DSR-E) | T(FRDM-FRDM) |
|---|---|---|---|---|---|---|---|---|
| At | 85,211 | 5.98 | 5.91 | 6.21E04 | 1.33E02 | 3.72E03 | 8.29E03 | 1.42E06 |
| At | 85,213 | 9.25 | 8.58 | 1.25E-07 | 6.00E-09 | 5.89E-08 | 7.82E-08 | 1.11E-02 |
| At | 85,215 | 8.18 | 8.67 | 1.00E-04 | 2.00E-05 | 3.31E-05 | 5.39E-05 | 6.13E-03 |
| At | 85,217 | 7.20 | 7.65 | 3.23E-02 | 4.00E-04 | 3.47E-02 | 6.95E-02 | 1.81E00 |
| Fr | 87,201 | 7.54 | 7.50 | 4.80E-02 | 1.50E-02 | - | 7.67E-02 | 5.02E01 |
| Fr | 87,207 | 6.90 | 6.49 | 1.56E01 | 3.44E-01 | 5.13E00 | 1.22E01 | 5.70E04 |
| Fr | 87,209 | 6.78 | 6.45 | 5.62E01 | 1.92E00 | 1.23E01 | 3.41E01 | 6.97E04 |
| Fr | 87,213 | 6.91 | 6.85 | 3.48E01 | 3.02E-01 | 3.80E00 | 9.03E00 | 2.28E03 |
| Fr | 87,215 | 9.54 | 9.38 | 8.60E-08 | 5.00E-09 | 5.50E-08 | 8.85E-08 | 5.71E-04 |
| Fr | 87,217 | 8.47 | 9.57 | 2.20E-05 | 5.00E-06 | 2.40E-05 | 5.00E-05 | 2.08E-04 |
| Fr | 87,219 | 7.45 | 7.38 | 2.00E-02 | 2.00E-03 | 3.02E-02 | 7.26E-02 | 3.53E01 |

| Parent | Z, A | Qα (exp) | Qα(FRDM) | T(E) | SE(T(E)) | T(T-E) | T(DSR-E) | T(FRDM-FRDM) |
|---|---|---|---|---|---|---|---|---|
| Fr | 87,221 | 6.46 | 6.09 | 2.94E02 | 1.20E01 | 1.58E02 | 4.25E02 | 7.38E05 |
| Ac | 89,215 | 7.75 | 7.50 | 1.70E-01 | 1.00E-02 | 2.34E-02 | 6.48E-02 | 7.84E01 |

**TABLE 2**. Atomic and mass numbers**,** the experimental and calculated Q value of alpha decay, the experimental and calculated values of the alpha decay half-lives, the standard error on the experimental alpha decay half-life, for the 225 TEST nuclides

Odd-Even

| Parent | Z, A | Qα (exp) | Qα(FRDM) | T(E) | SE(T(E)) | T(T-E) | T(DSR-E) | T(FRDM-FRDM) |
|---|---|---|---|---|---|---|---|---|
| Ac | 89,217 | 9.83 | 9.94 | 6.90E-08 | 4.00E-09 | 4.90E-08 | 9.74E-08 | 1.15E-04 |
| Ac | 89,219 | 8.83 | 9.89 | 1.18E-05 | 1.50E-06 | 1.23E-05 | 3.02E-05 | 1.26E-04 |
| Ac | 89,221 | 7.78 | 6.92 | 5.20E-02 | 2.00E-03 | 1.45E-02 | 4.00E-02 | 3.45E03 |
| Ac | 89,223 | 7.78 | 6.66 | 1.27E02 | 3.03E00 | 6.03E01 | 3.59E-02 | 2.35E04 |
| Pa | 91,215 | 8.17 | 7.91 | 1.40E-02 | 1.20E-02 | 6.46E-03 | 2.14E-02 | 2.01E01 |
| Pa | 91,217 | 8.49 | 8.06 | 4.90E-03 | 6.00E-04 | 6.17E-04 | 2.06E-03 | 6.90E00 |
| Pa | 91,219 | 10.08 | 10.51 | 5.30E-08 | 1.00E-08 | - | 1.36E-07 | 2.46E-05 |
| Pa | 91,221 | 9.25 | 10.70 | 5.90E-06 | 1.70E-06 | - | 1.34E-05 | 9.92E-06 |
| Pa | 91,223 | 8.34 | 7.10 | 6.50E-03 | 1.00E-03 | 1.41E-03 | 4.52E-03 | 3.28E03 |
| Pa | 91,225 | 7.39 | 7.62 | 1.70E00 | 2.00E-01 | 1.78E00 | 6.14E00 | 7.18E01 |
| Pa | 91,227 | 6.58 | 6.85 | 2.70E03 | 6.70E01 | 2.29E03 | 9.56E03 | 1.76E04 |

| Parent | Z, A | Qα (exp) | Qα(FRDM) | T(E) | SE(T(E)) | T(T-E) | T(DSR-E) | T(FRDM-FRDM) |
|---|---|---|---|---|---|---|---|---|
| Np | 93,227 | 7.82 | 8.14 | 5.10E-01 | 6.00E-02 | - | 1.40E00 | 8.04E00 |
| Am | 95,239 | 5.92 | 6.14 | 4.28E08 | 4.30E07 | 2.95E08 | 9.39E08 | 9.32E07 |

**TABLE 2**. Atomic and mass numbers**,** the experimental and calculated Q value of alpha decay, the experimental and calculated values of the alpha decay half-lives, the standard error on the experimental alpha decay half-life, for the 225 TEST nuclides

Odd-Even

| Parent | Z, A | Qα (exp) | Qα(FRDM) | T(E) | SE(T(E)) | T(T-E) | T(DSR-E) | T(FRDM-FRDM) |
|---|---|---|---|---|---|---|---|---|
| Es | 99,245 | 7.91 | 7.32 | 1.65E02 | 4.39E01 | 1.66E01 | 1.45E02 | 4.62E04 |
| Md | 101,255 | 7.91 | 8.02 | 2.03E04 | 5.28E03 | - | 7.63E02 | 5.81E02 |
| Md | 101,257 | 7.56 | 7.78 | 1.32E05 | 2.65E04 | - | 1.49E04 | 2.83E03 |
| Lr | 103,257 | 9.01 | 8.87 | 6.46E-01 | 2.50E-02 | - | 1.00E00 | 7.33E00 |

RMFDsq(E) = 0.23;   RMFDsq(T) = 0.74;   RMFDsq(DSR-E) = 0.98 ;   RMFDsq(FRDM-FRDM) = 7.10E04

**TABLE 2**. Atomic and mass numbers, the experimental and calculated Q value of alpha decay, the experimental and calculated values of the alpha decay half-lives, the standard error on the experimental alpha decay half-life, for the 225 TEST nuclides

Odd-Odd

| Parent | Z, A | $Q\alpha$ (exp) | $Q\alpha$(FRDM) | T(E) | SE(T(E)) | T(T-E) | T(DSR-E) | T(FRDM-FRDM) |
|---|---|---|---|---|---|---|---|---|
| Ta | 73,158 | 6.21 | 6.29 | 3.96E-02 | 3.08E-03 | 1.20E-02 | 1.91E-02 | 9.55E-02 |
| Ta | 73,160 | 5.55 | 5.28 | 4.41E00 | 5.88E-01 | - | 2.07E01 | 6.67E02 |
| Ta | 73,162 | 5.01 | 4.98 | 4.76E03 | 6.63E02 | - | 1.71E04 | 1.47E04 |
| Ir | 77,176 | 5.24 | 5.19 | 3.81E02 | 8.68E01 | - | 9.64E04 | 1.42E05 |
| Au | 79,182 | 5.53 | 6.15 | 5.53E04 | 1.19E04 | - | 2.57E04 | 1.11E02 |
| Bi | 83,190 | 6.86 | 7.06 | 7.68E00 | 2.35E00 | - | 9.24E-01 | 4.44E00 |
| At | 85,198 | 6.89 | 7.21 | 4.67E00 | 6.16E-01 | 1.12E00 | 4.63E00 | 9.05E00 |
| At | 85,200 | 6.60 | 6.91 | 7.54E01 | 8.13E00 | 1.23E01 | 8.50E01 | 8.30E01 |
| At | 85,202 | 6.35 | 6.67 | 3.23E02 | 2.44E02 | 3.02E02 | 1.06E03 | 5.38E02 |
| At | 85,204 | 6.07 | 6.20 | 1.45E04 | 8.27E02 | 1.95E03 | 2.49E04 | 3.07E04 |
| At | 85,210 | 5.63 | 5.42 | 1.67E07 | 2.07E06 | - | 4.55E06 | 7.29E07 |
| At | 85,212 | 7.83 | 7.57 | 3.14E-01 | 2.00E-03 | - | 4.71E-04 | 4.62E-01 |

| Parent | Z, A | Qα (exp) | Qα(FRDM) | T(E) | SE(T(E)) | T(T-E) | T(DSR-E) | T(FRDM-FRDM) |
|---|---|---|---|---|---|---|---|---|
| At | 85,214 | 8.99 | 8.78 | 5.58E-07 | 1.00E-08 | 2.34E-07 | 8.45E-08 | 2.68E-04 |

**TABLE 2**. Atomic and mass numbers**,** the experimental and calculated Q value of alpha decay, the experimental and calculated values of the alpha decay half-lives, the standard error on the experimental alpha decay half-life, for the 225 TEST nuclides

Odd-Odd

| Parent | Z, A | Qα (exp) | Qα(FRDM) | T(E) | SE(T(E)) | T(T-E) | T(DSR-E) | T(FRDM-FRDM) |
|---|---|---|---|---|---|---|---|---|
| At | 85,216 | 7.95 | 8.39 | 3.00E-04 | 3.00E-05 | 1.51E-04 | 1.44E-04 | 2.30E-03 |
| At | 85,218 | 6.87 | 6.70 | 1.60E00 | 4.00E-01 | 8.13E-01 | 1.82E00 | 2.62E02 |
| At | 85,220 | 6.05 | 5.64 | 2.80E03 | 7.07E02 | - | 1.26E04 | 4.90E06 |
| Fr | 87,206 | 6.93 | 6.55 | 1.81E01 | 8.52E-01 | 4.07E00 | 2.28E01 | 1.15E04 |
| Fr | 87,212 | 6.53 | 6.32 | 2.79E03 | 1.54E02 | - | 1.04E03 | 7.51E04 |
| Fr | 87,214 | 8.59 | 8.50 | 5.00E-03 | 2.00E-04 | - | 1.01E-05 | 8.24E-03 |
| Fr | 87,216 | 9.18 | 9.57 | 7.00E-07 | 2.00E-08 | 3.63E-07 | 1.56E-07 | 2.47E-05 |
| Fr | 87,218 | 8.01 | 8.26 | 1.00E-03 | 6.00E-04 | 5.01E-04 | 6.56E-04 | 3.10E-02 |
| Fr | 87,220 | 6.80 | 6.36 | 2.75E01 | 3.01E-01 | 6.76E00 | 3.74E01 | 4.12E04 |
| Ac | 89,218 | 9.38 | 9.83 | 1.12E-06 | 1.10E-07 | - | 2.56E-07 | 3.64E-05 |
| Ac | 89,220 | 8.35 | 7.77 | 2.61E-02 | 5.00E-04 | - | 3.46E-04 | 4.87E00 |
| Ac | 89,222 | 7.13 | 6.66 | 5.05E00 | 5.08E-01 | 2.29E00 | 1.31E01 | 2.42E04 |

| Pa | 91,218 | 9.79 | 9.61 | 1.20E-04 | 3.00E-05 | - | 1.26E-07 | 6.05E-04 |

**TABLE 2**. Atomic and mass numbers**,** the experimental and calculated Q value of alpha decay, the experimental and calculated values of the alpha decay half-lives, the standard error on the experimental alpha decay half-life, for the 225 TEST nuclides

Odd-Odd

---

| Parent | Z, A | $Q\alpha$ (exp) | $Q\alpha$(FRDM) | T(E) | SE(T(E)) | T(T-E) | T(DSR-E) | T(FRDM-FRDM) |
|---|---|---|---|---|---|---|---|---|
| Pa | 91,228 | 6.23 | 6.14 | 4.28E06 | 4.39E05 | - | 2.21E06 | 2.38E07 |
| Np | 93,236 | 5.02 | 4.99 | 3.04E15 | 7.69E14 | - | 1.05E15 | 1.71E14 |
| Bk | 97,244 | 6.78 | 6.29 | 2.61E08 | 8.75E07 | - | 2.36E06 | 2.90E09 |
| Md | 101,256 | 7.90 | 7.71 | 5.04E04 | 2.95E03 | - | 1.05E03 | 4.01E05 |
| Md | 101,258 | 7.27 | 7.06 | 4.45E06 | 2.59E04 | - | 5.55E05 | 9.16E07 |

---

RMFDsq(E) = 0.22;    RMFDsq(T) = 0.68;    RMFDsq(DSR-E) = 1.30 ;    RMFDsq(FRDM-FRDM) = 9.10E02

**TABLE 3.** Atomic and mass numbers, the experimental and calculated Q value of alpha decay, the experimental and calculated values of the alpha decay half-lives, for the 22 heavy and super-heavy elements (SHE).
See p. 25 for Explanation of Table.

| Parent | Z, A | $Q\alpha$ (exp) | $Q\alpha$(FRDM) | T(E) | T(R-E) | Even-Even T(DSR-E) | T(FRDM-FRDM) | T(DSR-E-fit) | T(FRDM-FRDM-fit) |
|---|---|---|---|---|---|---|---|---|---|
| Uub[a] | 112,282 | 9.53 | 9.42 | 7.20E00 | 5.38E00 | 5.18E00 | 4.20E01 | 2.37E00 | 3.12E-01 |
| Uub[a] | 112,284 | 9.14 | 8.70 | 3.63E01 | 8.10E01 | 6.96E01 | 1.90E03 | 3.44E01 | 2.31E01 |
| Uuq[a] | 114,286 | 10.00 | 9.40 | 7.00E-01 | 1.02E00 | 1.00E00 | 1.11E02 | 4.35E-01 | 9.31E-01 |
| Uuq[a] | 114,288 | 9.82 | 9.17 | 2.28E00 | 3.13E00 | 2.90E00 | 3.43E02 | 1.30E00 | 3.35E00 |
| Uus[a] | 116,290 | 10.77 | 11.12 | 1.05E-02 | 3.48E-02 | 3.70E-02 | 8.54E-02 | 1.45E-02 | 2.85E-04 |
| Uus[a] | 116,292 | 10.53 | 10.83 | 5.25E-02 | 1.38E-01 | 1.38E-01 | 2.65E-01 | 5.65E-02 | 1.02E-03 |
| Uuo[a] | 118,294 | 11.64 | 12.28 | 2.00E-03 | 9.66E-04 | 1.12E-03 | 1.95E-03 | 3.97E-04 | 4.00E-06 |

RMFDsq/R-E) = 1.21;   RMFDsq(DSR-E) = 1.22;  RMFDsq(FRDM-FRDM) = 84.28;  RMFDsq(DSR-E-fit) = 0.47;  RMFDsq(FRDM-FRDM-fit) = 0.78

**TABLE 3**. Atomic and mass numbers, the experimental and calculated Q value of alpha decay, the experimental and calculated values of the alpha decay half-lives, for the 22 heavy and super-heavy elements (SHE).

See p. 25 for Explanation of Table

Even-Odd

| Parent | Z, A | Qa (exp) | Qa(FRDM) | T(E) | T(R-E) | T(DSR-E) | T(FRDM-FRDM) | T(DSR-E-fit) | T(FRDM-FRDM-fit) |
|---|---|---|---|---|---|---|---|---|---|
| Hs | 108,267 | 10.11 | 9.35 | 2.60E-02 | 3.42E-02 | 2.87E-01 | 9.46E-01 | 1.53E-01 | 2.28E-03 |
| Uun[a] | 110,281 | 8.70 | 8.55 | 2.22E02 | 2.02E03 | 1.81E04 | 1.44E02 | 8.77E03 | 1.32E-01 |
| Uub[a] | 112,285 | 8.67 | 8.60 | 5.34E02 | 1.42E04 | 1.33E05 | 2.06E02 | 6.36E04 | 1.76E-01 |
| Uuq[a] | 114,289 | 9.71 | 8.87 | 6.70E01 | 3.30E01 | 3.33E02 | 6.42E01 | 1.67E02 | 6.85E-02 |

RMFDsq/R-E) = 13.41;  RMFDsq(DSR-E) = 130.91;  RMFDsq(FRDM-FRDM) = 17.69;  RMFDsq(DSR-E-fit) = 62.16; RMFDsq(FRDM-FRDM-fit) = 0.98

**TABLE 3.** Atomic and mass numbers, the experimental and calculated Q value of alpha decay, the experimental and calculated values of the alpha decay half-lives, for the 22 heavy and super-heavy elements (SHE).
See p. 25 for Explanation of Table.

Odd-Even

| Parent | Z, A | $Q\alpha$ (exp) | $Q\alpha$(FRDM) | T(E) | T(R-E) | T(DSR-E) | T(FRDM-FRDM) | T(DSR-E-fit) | T(FRDM-FRDM-fit) |
|---|---|---|---|---|---|---|---|---|---|
| Es | 99,241 | 8.32 | 7.97 | 9.00E00 | 4.06E00 | 6.48E00 | 5.08E02 | 1.57E01 | 4.35E00 |
| Md | 101,249 | 8.46 | 8.28 | 1.20E02 | 6.43E00 | 1.15E01 | 1.43E02 | 2.85E01 | 1.25E00 |
| Lr | 103,255 | 8.61 | 8.30 | 2.59E01 | 9.94E00 | 1.97E01 | 3.15E02 | 5.00E01 | 2.72E00 |
| Db | 105,263 | 8.83 | 8.28 | 6.28E01 | 8.85E00 | 1.94E01 | 8.16E02 | 4.94E01 | 6.91E00 |
| Bh | 107,261 | 10.56 | 10.33 | 1.24E-02 | 7.33E-04 | 1.50E-03 | 1.79E-02 | 2.54E-03 | 1.91E-04 |

RMFDsq/R-E) = 0.80;    RMFDsq(DSR-E) = 0.66;   RMFDsq(FRDM-FRDM) = 25.85;   RMFDsq(DSR-E-fit) = 0.73; RMFDsq(FRDM-FRDM-fit) = 0.87

**TABLE 3**. Atomic and mass numbers, the experimental and calculated Q value of alpha decay, the experimental and calculated values of the alpha decay half-lives, for the 22 heavy and super-heavy elements (SHE).

Odd-Odd

| Parent | Z, A | Qa (exp) | Qa(FRDM) | T(E) | T(R-E) | T(DSR-E) | T(FRDM-FRDM) | T(DSR-E-fit) | T(FRDM-FRDM-fit) |
|---|---|---|---|---|---|---|---|---|---|
| Es | 99,246 | 7.74 | 7.12 | 4.67e03 | 1.55E03 | 8.70E02 | 9.19E06 | 1.59E03 | 1.18E06 |
| Md | 101,248 | 8.70 | 8.47 | 3.50E01 | 3.95E00 | 1.26E00 | 1.86E03 | 4.18E00 | 2.00E02 |
| Lr | 103,254 | 8.85 | 8.49 | 1.67E01 | 7.14E00 | 2.11E00 | 9.02E03 | 6.64E00 | 1.00E03 |
| Db | 105,258 | 9.55 | 9.40 | 6.57E00 | 2.63E-01 | 5.38E-02 | 1.33E02 | 2.37E-01 | 1.36E01 |
| Bh | 107,264 | 9.97 | 9.65 | 4.40E-01 | 7.43E-02 | 1.22E-02 | 1.45E02 | 6.13E-02 | 1.49E01 |
| Uuu[b)] | 111,272 | 10.98 | 10.91 | 1.51E-03 | 3.33E-03 | 3.44E-04 | 3.13E00 | 2.40E-03 | 2.96E-01 |

RMFDsq/R-E) = 0.88;    RMFDsq(DSR-E) = 0.90;  RMFDsq(FRDM-FRDM) = 1.19E03;   RMFDsq(DSR-E-fit) = 0.77;

RMFDsq(FRDM-FRDM-fit) = 1.33E02

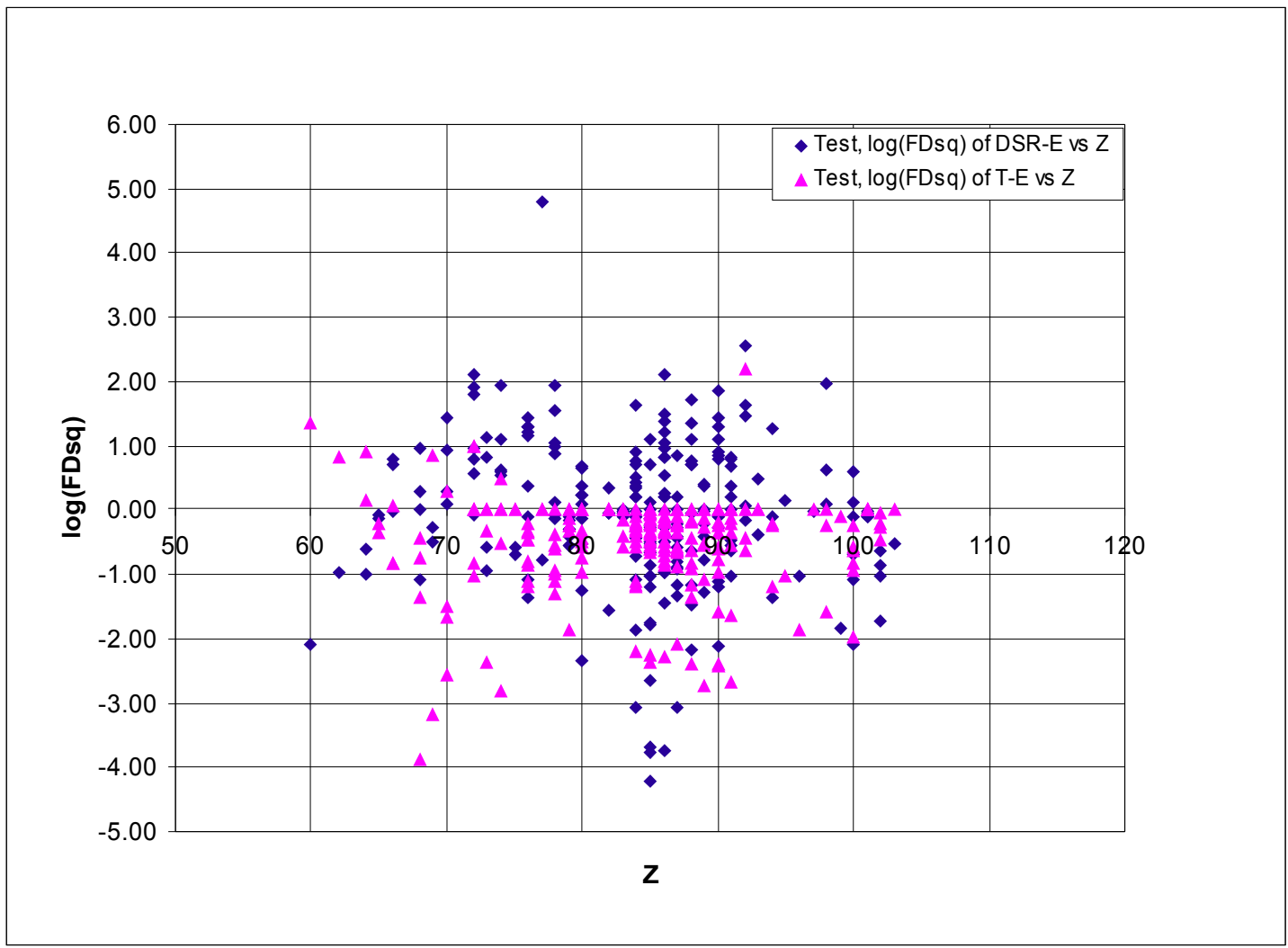

Fig. 1

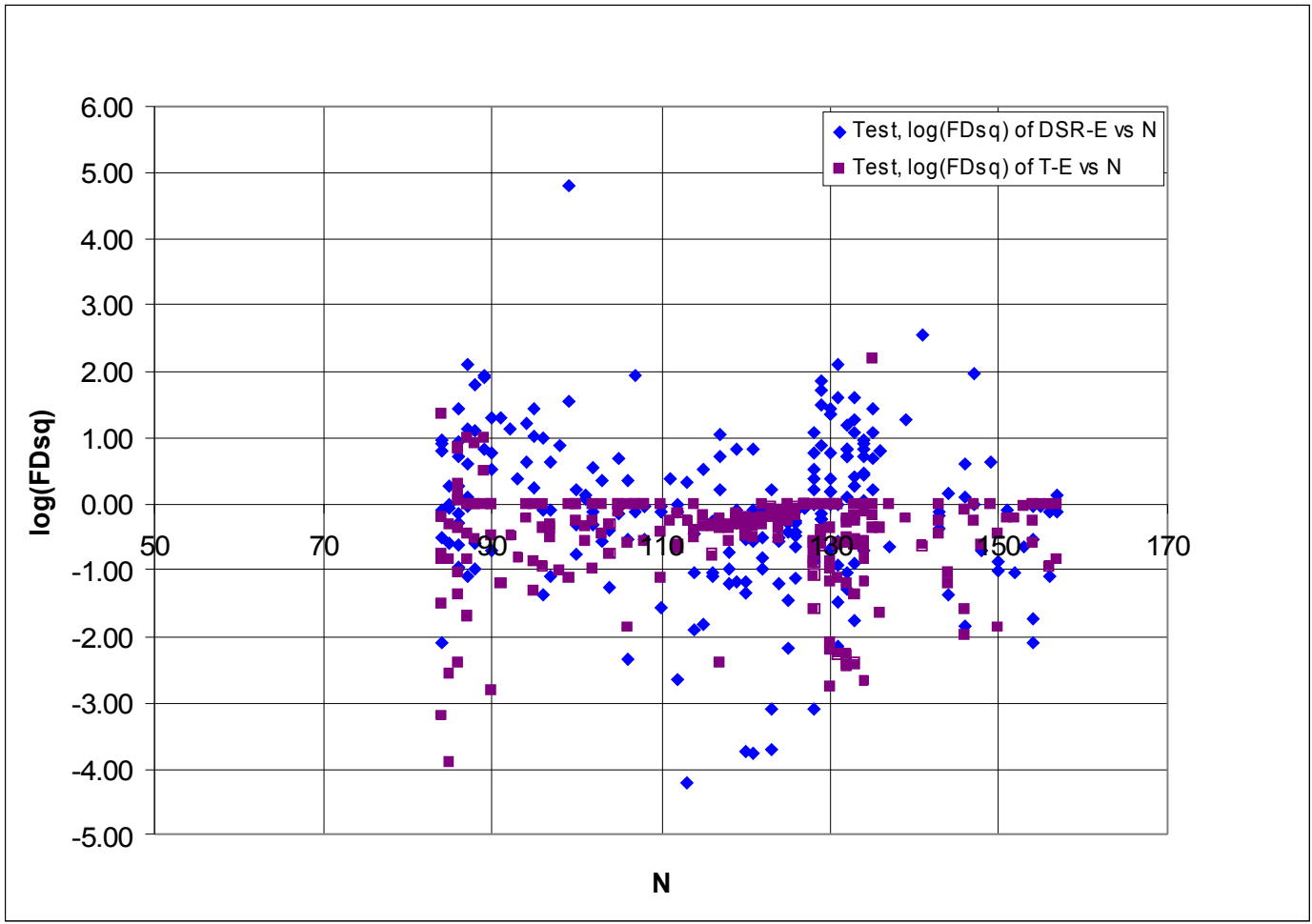

Fig.2

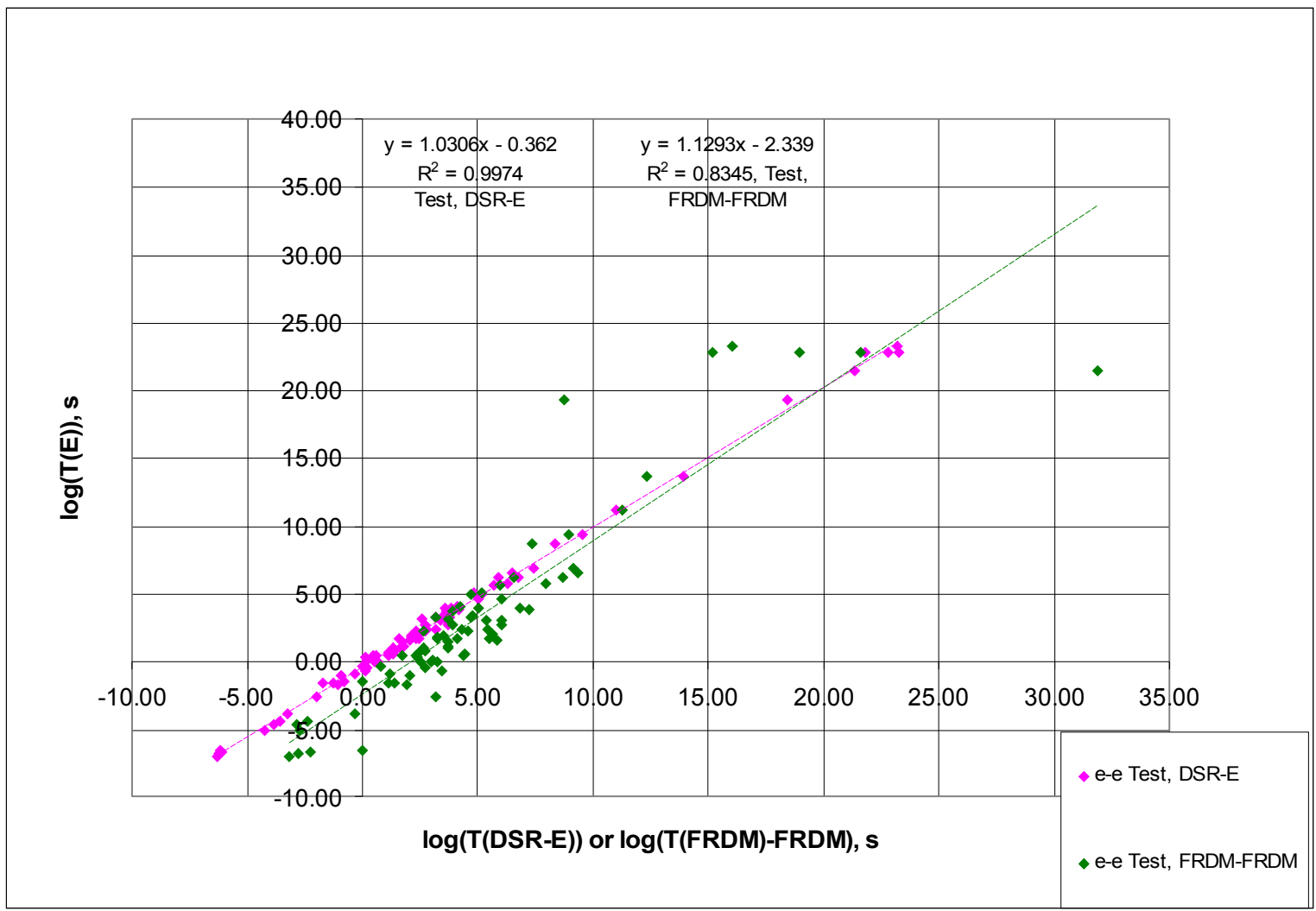

Fig. 3

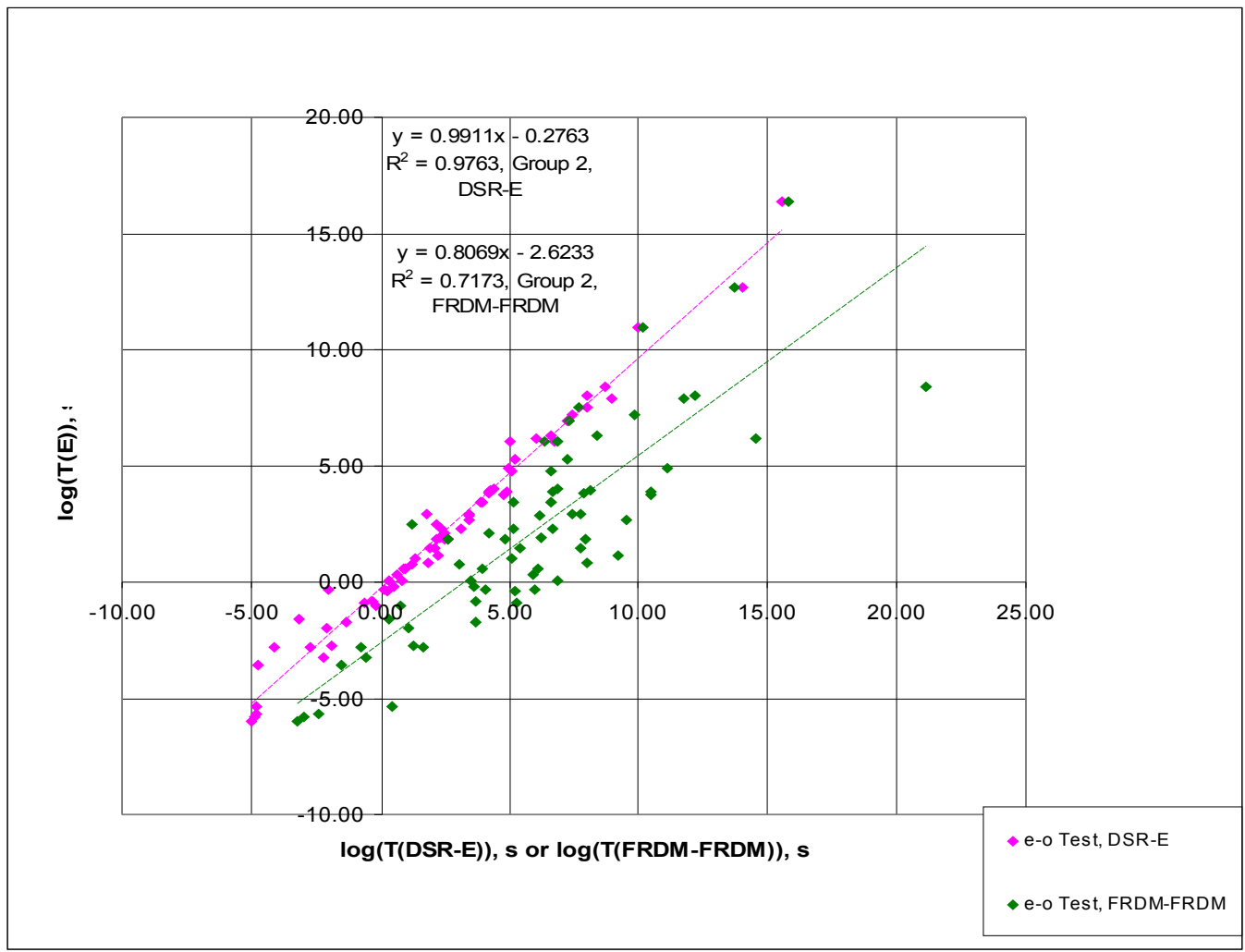

Fig. 4

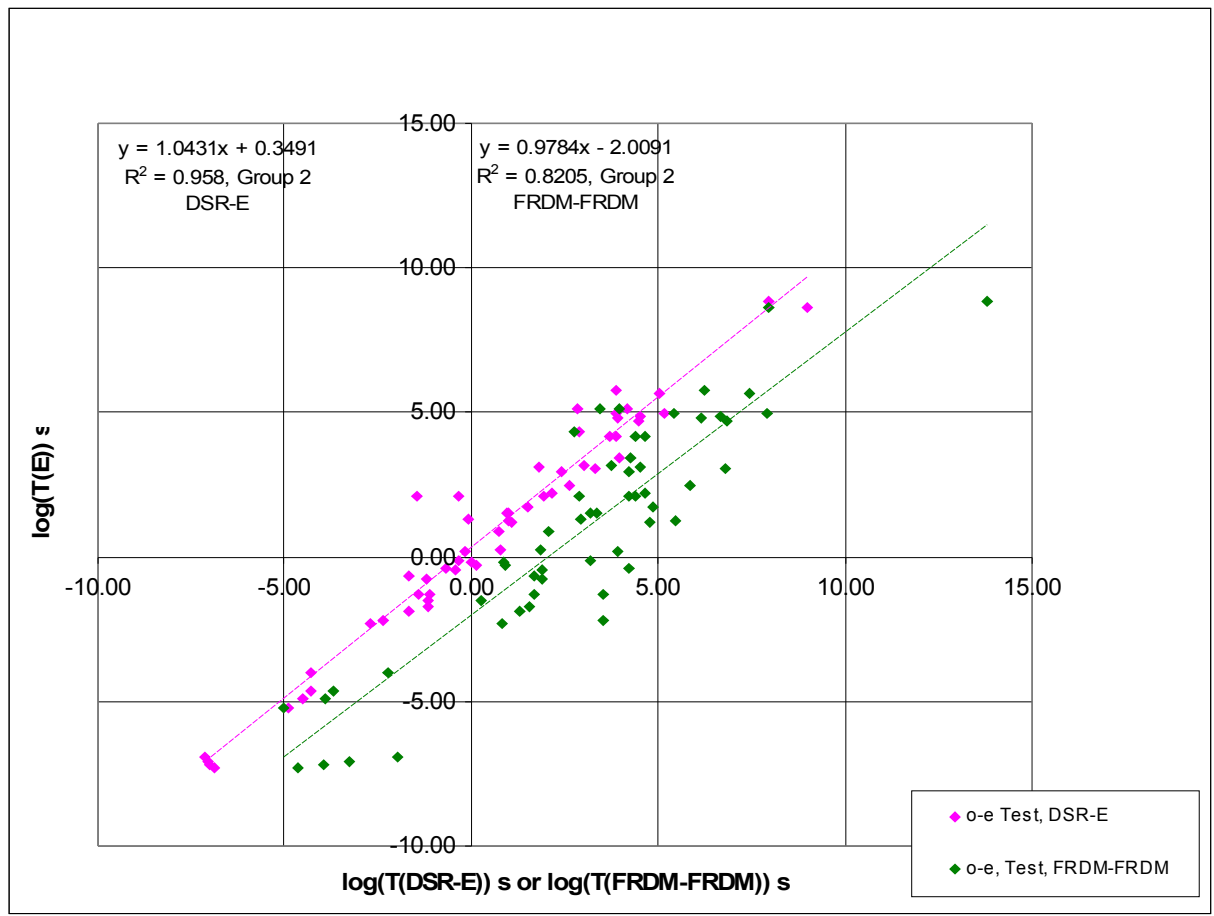

Fig. 5

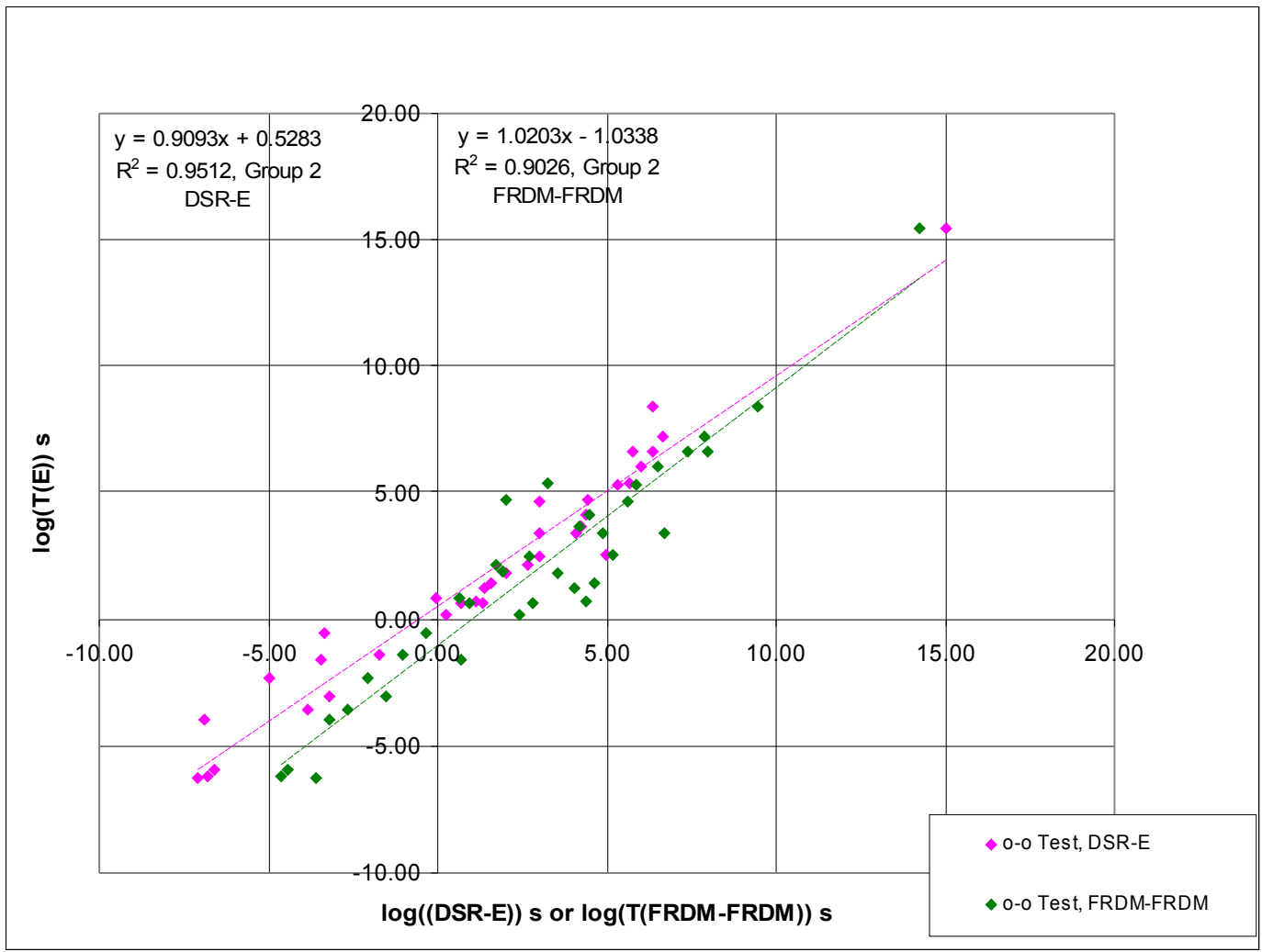

Fig. 6

**Figure Captions**

**Figure 1:** Variation of log(FDsq) with Z of the TEST nuclides the DSR and T calculations. Vide text for the meaning of the abbreviations.

**Figure 2:** Variation of log(FDsq) with N of the TEST nuclides for the DSR and T calculations. Vide text for the meaning of the abbreviations.

**Figure 3:** log(T(DSR-E)) and log(T(FRDM-FRDM)) vs log(T(E)) for even-even parent nuclides of TEST. Vide text for the meaning of the abbreviations.

**Figure 4:** log(T(DSR-E)) and log(T(FRDM-FRDM)) vs log(T(E)) for even-odd parent nuclides of TEST. Vide text for the meaning of the abbreviations.

**Figure 5:** log(T(DSR-E)) and log(T(FRDM-FRDM)) vs log(T(E)) for odd-even parent nuclides in TEST. Vide text for the meaning of the abbreviations.

**Figure 6:** log(T(DSR-E)) and log(T(FRDM-FRDM)) vs log(T(E)) for odd-odd parent nuclides of TEST. Vide text for the meaning of the abbreviations.